\documentclass[aps,twocolumn,floatfix,altaffilletter,superscriptaddress,preprintnumbers,tightenlines,showpacs,showkeys,notitlepage,nofootinbib]{revtex4-2}

\usepackage[T1]{fontenc}
\usepackage[colorlinks=true,citecolor=blue,linkcolor=blue]{hyperref}
\usepackage{braket,graphicx,physics,amsthm,amssymb,amsfonts,xcolor,booktabs,siunitx,mathrsfs}
\usepackage{mathtools}
\usepackage{slashed}
\usepackage{bm}
\usepackage[capitalize]{cleveref}
\usepackage{titlesec}
\usepackage[caption=false]{subfig}
\usepackage{multirow}

\usepackage[normalem]{ulem}
\usepackage{slashed}	
\usepackage{gensymb}
\usepackage{soul}

\allowdisplaybreaks  

\makeatletter

\newif\ifhidetocsections
\hidetocsectionsfalse

\newif\ifhidetocsubsections
\hidetocsubsectionsfalse

\let\oldl@section\l@section
\let\oldl@subsection\l@subsection
\let\oldl@subsubsection\l@subsubsection

\renewcommand{\l@section}[2]{%
  \ifhidetocsections\else\oldl@section{#1}{#2}\fi}

\renewcommand{\l@subsection}[2]{%
  \ifhidetocsubsections\else\oldl@subsection{#1}{#2}\fi}

\renewcommand{\l@subsubsection}[2]{%
  \ifhidetocsubsections\else\oldl@subsubsection{#1}{#2}\fi}
\makeatother

\providecommand*{\diff}%
	{\@ifnextchar^{\DIfF}{\DIfF^{}}}
\def\DIfF^#1{%
	\mathop{\mathrm{\mathstrut d}}%
		\nolimits^{#1}\gobblespace}
\def\gobblespace{%
	\futurelet\diffarg\opspace}
\def\opspace{%
	\let\DiffSpace\!%
	\ifx\diffarg(%
		\let\DiffSpace\relax
	\else
		\ifx\diffarg[%
			\let\DiffSpace\relax
		\else
  			\ifx\diffarg\{%
				\let\DiffSpace\relax
			\fi\fi\fi\DiffSpace}

\usepackage{xspace}

\definecolor{cbred}{HTML}{d55e00}
\definecolor{cborange}{HTML}{e69f00}
\definecolor{cbgreen}{HTML}{009e73}
\definecolor{cbyellow}{HTML}{f1dd42}
\definecolor{cblblue}{HTML}{56b4e9}
\definecolor{cbblue}{HTML}{0072b2}
\definecolor{cbpink}{HTML}{cc79a7}
\definecolor{defgrey}{HTML}{9f9f9f}
\definecolor{defgreen}{HTML}{8eba42}

\makeatletter
\renewcommand\onecolumngrid{%
  \do@columngrid{one}{\@ne}%
  \def\set@footnotewidth{\onecolumngrid}%
  \def\footnoterule{\kern-6pt\hrule width 1.5in\kern6pt}%
}
\makeatother

\begin{document}

\preprint{SISSA 04/2026/FISI}
\title{Vector Resonances at Muon and Wakefield Colliders}

\author{Massimo Cipressi}
\affiliation{SISSA International School for Advanced Studies, Via Bonomea 265, 34136, Trieste, Italy}
\affiliation{INFN, Sezione di Trieste, Via Bonomea 265, 34136, Trieste, Italy}

\author{Kevin Langhoff}
\affiliation{Center for Theoretical Physics -- a Leinweber Institute, Massachusetts Institute of Technology,\\
77 Massachusetts Avenue, Cambridge, Massachusetts, USA}

\author{Toby Opferkuch}
\affiliation{SISSA International School for Advanced Studies, Via Bonomea 265, 34136, Trieste, Italy}
\affiliation{INFN, Sezione di Trieste, Via Bonomea 265, 34136, Trieste, Italy}

\begin{abstract}
We explore the potential of future high-energy lepton colliders to probe heavy vector resonances. At wakefield colliders, intense beam-beam interactions produce radiation, called beamstrahlung, which redistributes luminosity from the nominal energy across a broad spectrum of lower collision energies. We show that this effect, conventionally viewed as a drawback, dramatically enhances sensitivity to resonances by effectively scanning a wide range of center-of-mass energies. We present projections for a benchmark scenario of a heavy kinetically mixed $Z'$. 
\end{abstract}


\maketitle

\section{Introduction}
\label{sec:Introduction}

As the Large Hadron Collider (LHC) enters its precision era, the community is simultaneously shaping the long-term energy-frontier strategy. 
Candidate paths include a higher-energy hadron machine (FCC-hh), a muon collider (MuC), and a \emph{wakefield collider} (WFC)~\cite{Adolphsen:2022ibf, P5:2023wyd, Gessner:2025acq}.

A collider's sensitivity to new heavy physics depends not only on its energy and luminosity, but also on how it probes energies below the nominal center-of-mass collision energy,  $\sqrt{s}$. 
Hadron colliders exploit the composite nature of protons to provide continuous coverage of partonic energies. 
A MuC accesses lower energies by emitting collinear-enhanced initial-state radiation (ISR) \cite{Chakrabarty:2014pja,Karliner:2015tga}. 
A WFC has strong beam-beam interactions at bunch crossing which generate new particles and redistribute luminosity across a broad energy spectrum via \emph{beamstrahlung}~\cite{YokoyaChen:1992, Esberg:2014zia} (in the quantum regime of multi-TeV designs these spectra take a universal form~\cite{He:2025jmz}). 

\begin{figure}[t]
    \centering
    \includegraphics[width=1.01\linewidth]{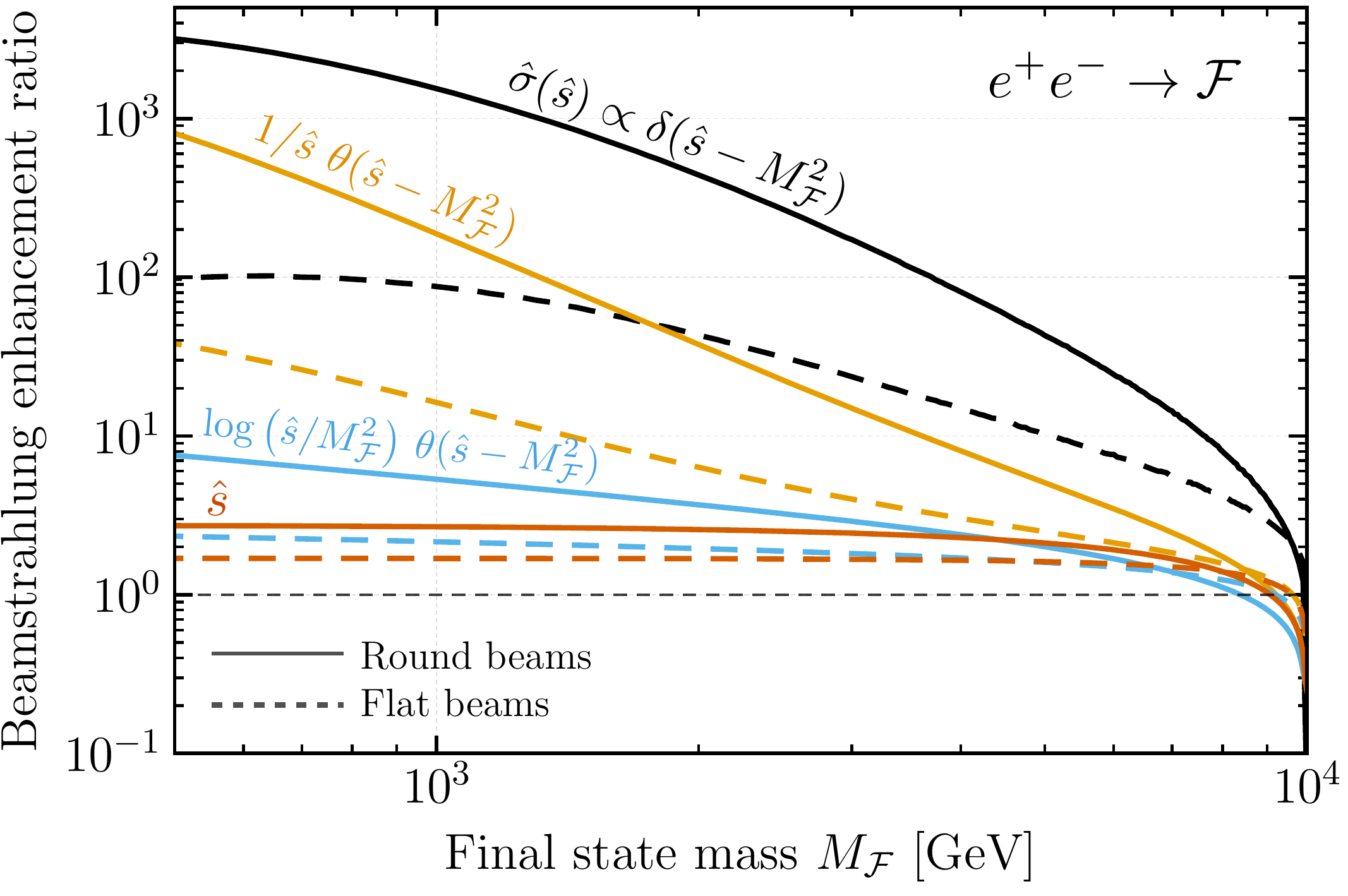}
    \caption{Ratio of $e^+e^-$ cross sections with and without beamstrahlung as a function of the final-state threshold energy $M_{\mathcal{F}}$, for representative parton-level cross section scalings: resonance production ($\delta$-function), $s$-channel ($\hat{s}^{-1}$), vector boson fusion ($\log\hat{s}$), and dimension-6 operators ($\hat{s}$); see \cref{eq:R_ij,eq:repr-xsecs}. Solid (dashed) lines correspond to round (flat) wakefield collider beams.\vspace{-0.5cm}}
    \label{fig:Rratio}
\end{figure}

We show that beamstrahlung, conventionally viewed as a drawback, gives a WFC a significant advantage for resonance searches. We emphasize that ISR at any lepton collider already provides access to sub-nominal energies; our MuC projections fully account for this. The advantage of beamstrahlung is quantitative: at a WFC the luminosity at sub-nominal energies exceeds the ISR luminosity by orders of magnitude and this enhancement drives the improved sensitivity to weakly coupled resonances. The enhancement from beamstrahlung relative to ISR alone for resonance production (black lines) and various other representative cross sections are shown in \cref{fig:Rratio}; details are explained later in \cref{eq:R_ij} and \cref{eq:repr-xsecs}. 

Beamstrahlung also offers a partial remedy to the positron acceleration problem of WFCs~\cite{PhysRevE.64.045501,Hogan:2003bs,PhysRevAccelBeams.27.034801}, potentially limiting a WFC to electron-only beams. Positrons and photons produced via beamstrahlung provide access to collision channels that would otherwise be unavailable.

As a benchmark,\footnote{Our projections can be recast for any narrow resonance coupling to leptons via a simple rescaling of the signal strength $\mu_{Z'}$. 
The relative advantage of different colliders is model-dependent; for example, $L_i-L_j$ gauge bosons are elusive at the LHC~\cite{Dasgupta:2023zrh}.}
we consider a $Z'$ coupled to the Standard Model (SM) through kinetic mixing \cite{Okun:1982xi,Galison:1983pa,Holdom:1985ag,Babu:1997st,Dienes:1996zr,Langacker:2008yv,Fabbrichesi:2020wbt} and compare projections for a MuC \cite{Azatov:2022itm,Airen:2024iiy,Cheung:2025uaz} and several WFC configurations with $\sqrt{s} = \SI{10}{\TeV}$ against existing and future hadron colliders \cite{Curtin:2014cca,FCC_hh,FCC:2018byv}.  
At fixed $\sqrt{s}$ and geometric luminosity, an $e^+e^-$ WFC improves sensitivity to the kinetic mixing parameter $\varepsilon$ by more than an order of magnitude compared to a MuC for $M_{Z^\prime} = \SI{1}{\TeV}$. 
Even without a primary positron beam, electron-only and photon WFC configurations benefit significantly from secondary particles produced via beamstrahlung.

We describe WFC features that differ from conventional lepton colliders in \cref{sec:Wakefield Accelerators and Colliders}.
In \cref{sec:Resonance Searches} we develop the resonance search strategy in a unified formalism applicable to a MuC and WFC, present numerical projections, and compare across collider options.  We conclude in \cref{sec:Conclusion}. 
Details about analytic approximations and the analysis are collected in \cref{app:analytic-approximation-details,app:analysis-details}.
\section{Wakefield Colliders} 
\label{sec:Wakefield Accelerators and Colliders}
Wakefield acceleration leverages large electric fields, potentially exceeding \SI{100}{\giga\volt\per\meter}, excited by driver beams in a plasma medium~\cite{PhysRevLett.43.267, PhysRevLett.54.693, Esarey:2009}. 
These surpass RF cavity gradients by orders of magnitude, offering a compact path to multi-TeV energies. 
While significant technical challenges remain~\cite{Gessner:2025acq, Schroeder_2022}, this work focuses on the resulting phenomenology: 
specifically, how high energies and charge densities of wakefield beams alter the collision environment through beamstrahlung and how this can be exploited in searches for new physics.
\begin{figure*}
    \centering
    \includegraphics[height=0.43\linewidth]{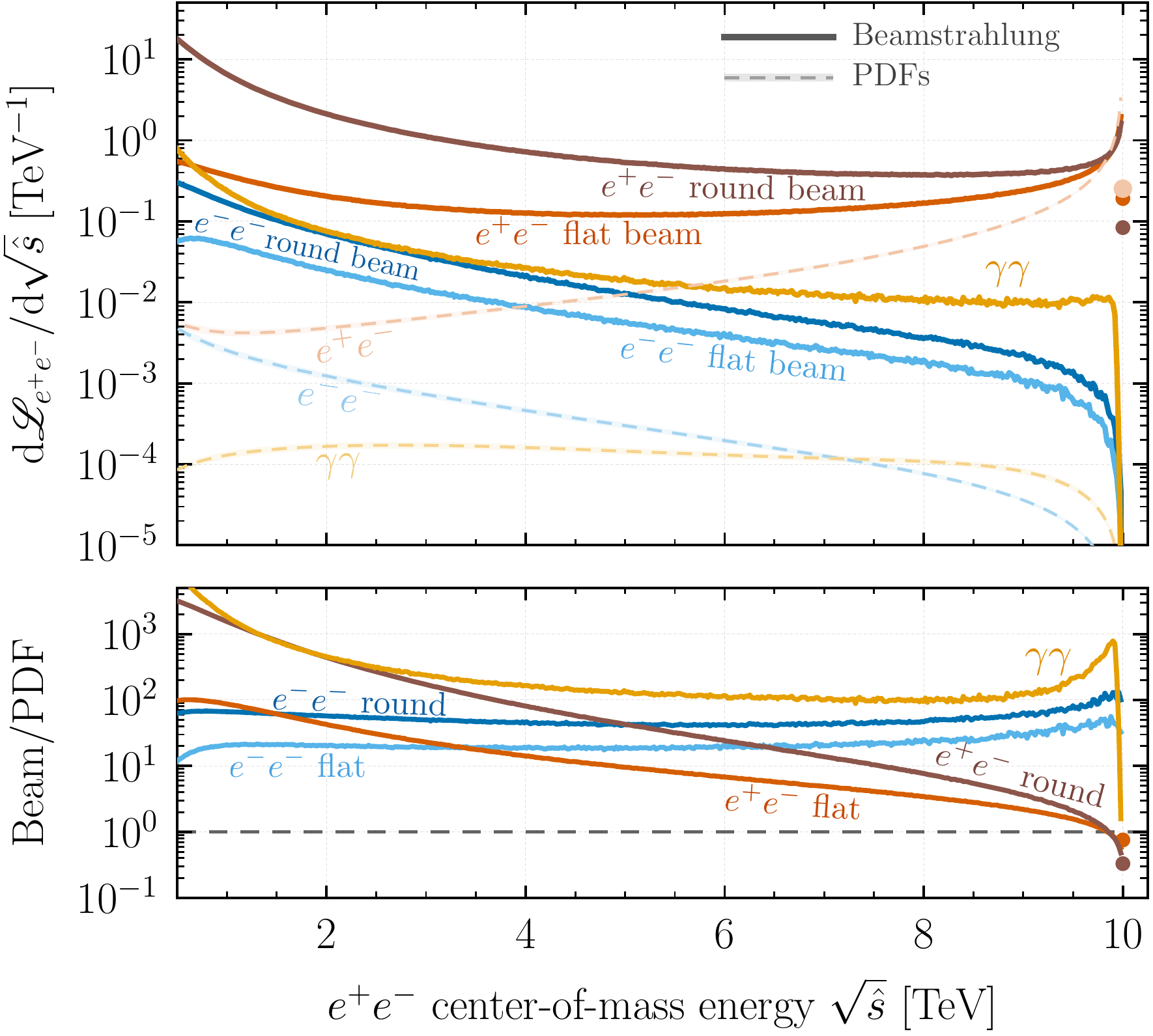}
    \raisebox{0.6ex}{\includegraphics[height=0.426\linewidth]{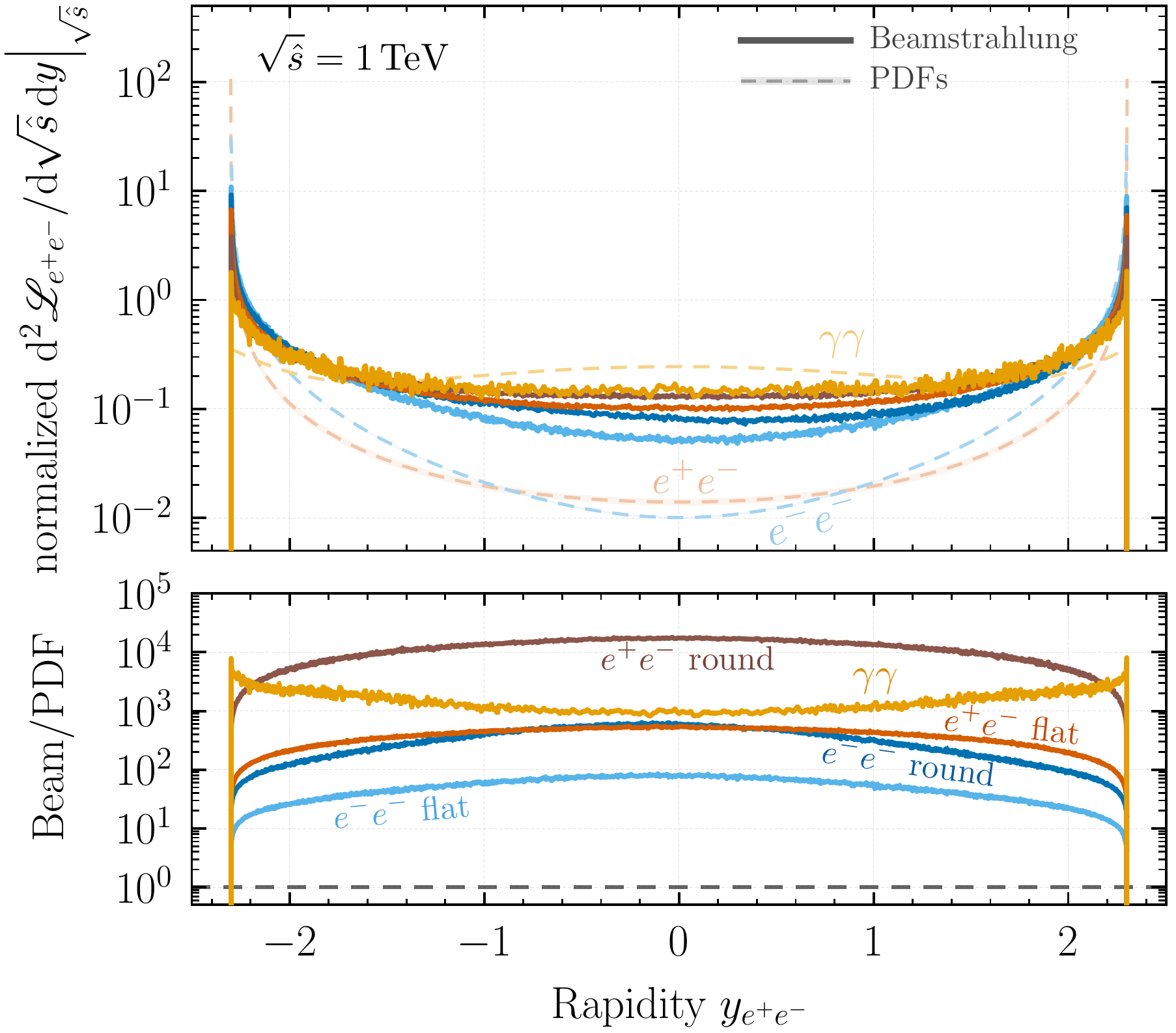}}
    \caption{\textbf{(Left)} Differential luminosity as a function of partonic center-of-mass energy $\sqrt{\hat{s}}$ for $e^+e^-$ initial states and $\sqrt{s} = \SI{10}{\TeV}$. Thick curves are simulation results~\cite{acceleratorpaper,simulationpaper} incorporating beamstrahlung effects. For comparison, thin dashed lines show the PDF-only contributions from ISR in the absence of beamstrahlung. The lower panel shows their ratio. The round markers at $\sqrt{\hat{s}}=\SI{10}{\TeV}$ indicate the luminosity integrated over the last $\SI{5}{\GeV}$ which we treat as a delta-function. \textbf{(Right)} Normalized double-differential luminosity as a function of the $e^+e^-$ rapidity for $\sqrt{\hat s} = \SI{1}{\TeV}$. The lower panel shows the ratio of the unnormalized beamstrahlung and PDF-only contributions.
    } 
    \label{fig:beam-beam-vs-machine}
\end{figure*}
\subsection{Beamstrahlung} 
\label{subsec:Beamstrahlung}
High luminosity at a multi-TeV WFC requires transverse beam sizes $\sigma_{x,y} \simeq \mathcal{O}(\SI{1}{\nano\meter})$. 
The bunch length is also constrained to $\sigma_z\lesssim \mathcal{O}(\SI{10}{\micro\meter})$ to fit into the plasma wake. 
High energies and charge densities produce intense electromagnetic fields during the bunch crossing, characterized by the beamstrahlung parameter~\cite{Schroeder_2022, Esberg:2014zia}
\begin{equation}
  \Upsilon = \frac{5}{6}
    \frac{\alpha \gamma\,N}{m_e^2\,\sigma_z\,(\sigma_x+\sigma_y)}\,,
\end{equation}
where $\gamma$ is the Lorentz factor, $N$ the particles/bunch. 
For \SI{10}{\TeV} WFC designs, $\Upsilon$ can exceed $\mathcal{O}(10^3)$, well into the quantum beamstrahlung regime~\cite{Roser:2022sht}.

The integrated geometric luminosity
\begin{align}
  L_{\rm geom} = \frac{N^2 f\, T}{4\pi \sigma_x\,\sigma_y}\,,
\end{align}
scales as $(\sigma_x \sigma_y)^{-1}$, where $f$ is the bunch crossing frequency and $T$ the total integrated time. 
For fixed $\sigma_x \sigma_y$ and therefore geometric luminosity, ``flat beams'' ($\sigma_x \gg \sigma_y$) reduce $\Upsilon$ relative to ``round beams'' ($\sigma_x = \sigma_y$). 
However, flat beams at wakefield accelerators are subject to nonlinear perturbations which trigger transverse emittance mixing and degrade luminosity~\cite{Diederichs:2024zir}. 
The \SI{10}{\TeV} WFC design study is therefore exploring both flat and round beam configurations~\cite{Gessner:2025acq}. 

A further consideration is the choice of beam particles. Accelerating positrons in a plasma wake is significantly more difficult than accelerating electrons because the plasma response is not charge-symmetric~\cite{PhysRevAccelBeams.27.034801}. If high-quality positron beams are unavailable, one may either use an $e^-e^-$ collider or convert one or both electron beams to photons via Compton backscattering to realize a $\gamma\gamma$ collider~\cite{Ginzburg:1983,Telnov:1990}. We compare several configurations of beam species and geometries in what follows.
\subsection{Luminosity Spectra}
\label{subsec:Luminosity Spectra}
For all lepton collider configurations in this work we fix the integrated geometric luminosity to $L_{\rm geom} = \SI{10}{\atto\barn^{-1}}$.  
This quantity is determined by the beam parameters alone---bunch charge, crossing rate, and transverse spot sizes---and does not account for the beamstrahlung effects described above. 
To incorporate these effects, we introduce \emph{luminosity spectra}: dimensionless distributions $d\mathscr{L}_{ij}/d\tau\,dy$ describing collisions of species $i$ and $j$ per unit geometric luminosity as a function of $\tau = \hat{s}/s$ and rapidity $y$, where $\sqrt{\hat{s}}$ denotes the collision center-of-mass energy. 
This notation is in direct analogy with parton luminosity functions, $d\mathcal{L}_{ij}/d\tau\,dy$. 
The analogy is not exact: for wakefield colliders, colliding particles are on-shell states produced by beam-beam interactions, while parton luminosity functions describe distributions of slightly off-shell partons. 
Despite this distinction, these enter cross sections identically with the mapping\footnote{Strictly speaking, parton luminosity functions should also be convolved with the beamstrahlung luminosity spectra to account for additional off-shell collinear radiation from the secondary particles. 
This is a sub-leading correction for $\tau \ll 1$, where the beamstrahlung spectra dominate, but becomes relevant near $\tau \to 1$.}
\begin{align} \label{eq:L_to_L_Relation}
  \frac{d\mathcal{L}_{ij}}{d\tau\,dy}
  \;\longleftrightarrow\;
  \frac{d\mathscr{L}_{ij}}{d\tau\,dy}\,.
\end{align}
Luminosity spectra used in this work are obtained from simulations of bunch-bunch interactions~\cite{acceleratorpaper,simulationpaper} performed with \texttt{WarpX}~\cite{warpx} ($e^+e^-$ and $e^-e^-$ configurations) and \texttt{CAIN}~\cite{Chen:1994jt} ($\gamma\gamma$). 
These use beam parameters developed by the LBNL BELLA group~\cite{acceleratorpaper}. 
All configurations assume $N = 1.2\times 10^9$ particles/bunch and $\sigma_z = \SI{8.5}{\micro\meter}$. ``Flat beams'' have $(\sigma_x,\sigma_y) = (6,\,0.4)\,\SI{}{\nano\meter}$ and ``round beams'' have $(\sigma_x,\sigma_y) = (1.55,\,1.55)\,\SI{}{\nano\meter}$. Further details are given in Table~1 of \cite{PWFA}.

Luminosity spectra are shown in comparison to electroweak parton luminosity functions (PDFs) for $e^+e^-$ collisions in \cref{fig:beam-beam-vs-machine} (see \cref{fig:beam-beam-vs-machine-gg} for $\gamma\gamma$ collisions). 
The PDF-only contributions from ISR ($d\mathcal{L}_{e^+e^-}/d\tau$) are computed using LePDF~\cite{Garosi:2023bvq}; see \cref{app:analytic-approximation-details} for additional details. 
For most energies, $d\mathscr{L}_{e^+e^-}/d\tau$ exceeds $d\mathcal{L}_{e^+e^-}/d\tau$ by orders of magnitude, confirming beamstrahlung dominates over ISR. 
Beamstrahlung also produces broader rapidity distributions than those from ISR alone.

To quantify the effects of beamstrahlung we define the ratio of cross sections with and without beamstrahlung as
\begin{align} \label{eq:R_ij}
    R_{ij}(M_{\mathcal{F}},s) = \frac{\int_{\tau_\mathcal{F}}^1 d\tau \, (d\mathscr{L}_{ij}/d\tau)\, \hat{\sigma}(\tau s)  }{\int_{\tau_\mathcal{F}}^1 d\tau \, (d\mathcal{L}_{ij}/d\tau)\, \hat{\sigma}(\tau s) }\,.
\end{align} 
In this ratio, we consider the production of a final state $\mathcal{F}$ with threshold energy $M_{\mathcal{F}}$ and representative parton-level cross sections:
\begin{align}\label{eq:repr-xsecs}
    \hat{\sigma}(\hat{s}) \propto 
    \begin{cases}
        \delta(\hat{s} - M_{\mathcal{F}}^2), & (\text{Resonance})\\
        \hat{s}^{-1}\,\theta(\hat{s} - M_{\mathcal{F}}^2), & (s\text{-channel})\\
        \log\left(\hat{s}/ M_{\mathcal{F}}^2\right)\theta(\hat{s} - M_{\mathcal{F}}^2), & (\text{VBF})\\
        \hat{s}, & (\text{Dim-6})
    \end{cases}
\end{align}
This ratio is shown for $e^+e^-$ initial states at $\sqrt{s} = \SI{10}{\TeV}$ in \cref{fig:Rratio}. 
The enhancement is largest for cross sections weighted towards the lowest energies: both resonance and $s$-channel production rates are boosted by orders of magnitude.
This motivates dedicated searches for resonances, which we pursue below, as well as studies of $s$-channel pair production of heavy electroweak states~\cite{PWFA}.
\section{Resonance Searches at Lepton Colliders} 
\label{sec:Resonance Searches}
We consider lepton collider searches for a heavy $Z'$ coupled to the SM only through kinetic mixing
\begin{align}
    \mathcal{L} \supset \frac{\varepsilon}{2\cos\theta_W}\, B^{\mu\nu} Z'_{\mu\nu}\,.
\end{align}
Upon diagonalization of the kinetic terms, couplings to SM fermions are proportional to the hypercharge current
\begin{align}
    \mathcal{L} \supset \frac{\varepsilon \, e}{\cos^2\theta_W}  Z'_\mu J_Y^\mu,\quad  {\rm for}~M_{Z'} \gg M_Z\,.
\end{align}
We assume the $Z^\prime$ decays only to SM final states. The total width is $\Gamma_{Z'} = 5\alpha\varepsilon^2 M_{Z'}/3\cos^4\theta_W$ and the branching ratio to each lepton flavor $\ell$ is ${\rm BR}(Z' \to \ell^+\ell^-) \approx 1/8$. 

The formalism for resonance searches at a MuC and WFC is similar. 
At a MuC, collinear ISR reduces the lepton collision energy, and the resulting $\mu^+\mu^-$ collision 
energies are described by the parton luminosity function $d\mathcal{L}_{\mu^+ \mu^-}/d\tau\, dy$. At a WFC, beamstrahlung plays an analogous role and $\mathcal{L}$ is substituted with $\mathscr{L}$ when computing WFC cross sections as in \cref{eq:L_to_L_Relation}.

We consider two search strategies for di-lepton resonances. 
In the \emph{inclusive} search
\begin{align}
    \ell^+\ell^- \to ( Z' \to \ell^{\prime +}\ell^{\prime -}) + X\,,
\end{align}
the ISR photon that reduces the collision energy is unresolved---it escapes down the beam pipe---and its effect is captured by PDFs (or, at a WFC through beamstrahlung luminosity spectra). 
In the \emph{exclusive} search 
\begin{align}
    \ell^+\ell^- \to (Z^\prime \to \ell^{\prime +}\ell^{\prime -})  +\gamma
\end{align}
the ISR photon is hard and central enough to be reconstructed in the detector; this requires retaining the angular distribution of the emitted photon, as the collinear PDF approximation does not capture finite-angle emission.
In both cases the flavor of $\ell'$ differs from $\ell$ to remove $t$-channel backgrounds, e.g. $e^+e^- \rightarrow Z'\rightarrow \mu^+ \mu^-$.

At a MuC, both searches play important roles: 
the inclusive search dominates at high masses, while the exclusive search recovers sensitivity at low masses where the boosted di-lepton system falls outside the detector acceptance. 
At a WFC, the broad rapidity distribution from beamstrahlung keeps signal events sufficiently central across the full mass range, making the exclusive search subdominant; we therefore present only the inclusive analysis for WFC configurations. 
We first motivate the analysis using analytic approximations, highlighting parametric scaling and kinematic features.

\subsection{Analytic Approximations}

In this subsection, we focus on the inclusive search, where ISR is collinear and well described by parton distribution functions; explicit expressions for the lepton and photon luminosity functions used below are collected in \cref{app:analytic-approximation-details}. 
Analytic approximations for the exclusive search, which requires retaining the angular distribution of the ISR photon, are given in \cref{sec:Exclusive Resonance Production}.

\emph{Signal:} We assume $\Gamma_{Z'}\ll M_{Z'}$ and use the \emph{narrow-width approximation}. 
The total cross section is proportional to the luminosity at energy $M_{Z'}$
\begin{align}\label{eq:sigma_X}
    \sigma_{Z'}(s) \approx 12\pi^2 \mu_{Z'}  \frac{1}{s} \frac{d\mathcal{L}_{\ell \ell}}{d\tau} \Big|_{\tau = \tau_{Z'}}\,, 
\end{align}
where $\tau_{Z'} = M_{Z'}^2/s$ and we define the signal strength
\begin{align}
    \mu_{Z'} \equiv \frac{\Gamma(Z'\rightarrow \ell \ell)\,{\rm BR}(Z'\rightarrow \ell' \ell')}{M_{Z'}}\,.
\end{align}
For the kinetically mixed $Z'$ with $M_{Z'}\gg M_Z$ we have $\mu_{Z'}\approx (5/192)\alpha \varepsilon^2/(\cos^4\theta_W)\approx 3.43\times 10^{-4}\, \varepsilon^2$.

For both a MuC and a WFC, the signal is maximized for collisions involving one lepton with near-maximal energy; 
therefore, the rapidity distribution of the resonance peaks near $|y_{Z'}| = \log(\sqrt{s}/M_{Z'})$. 
However, the rapidity distribution at a WFC is broader than the sharply peaked distribution characteristic of ISR at a MuC, as seen in the luminosity spectra in \cref{fig:beam-beam-vs-machine} and the event-level rapidity distributions in \cref{fig:Rapidity_Comparison_Combined}. 
These features motivate selection cuts on the di-lepton 
invariant mass and rapidity:
\begin{align}
    &\Big|m_{\ell' \ell'} - M_{Z'}\Big| < \Delta \, M_{Z'}\,, \label{eq:dilepton_mass_cut}\\
    &\Big||y_{\ell' \ell'}| - y_{Z'}\Big| < \delta\,. \label{eq:dilepton_rapidity_cut}
\end{align}
Values of $\Delta$ and $\delta$ will be motivated from lepton energy and $p_T$ resolutions. 
We first consider backgrounds after only the invariant mass cut, then turn to (pseudo)rapidity acceptance.\\

\begin{figure*}[ht!]
    \centering
    \begin{minipage}[t]{0.49\linewidth}
        \centering
        \includegraphics[width=\linewidth,trim=0 0 0 4pt,
    clip]{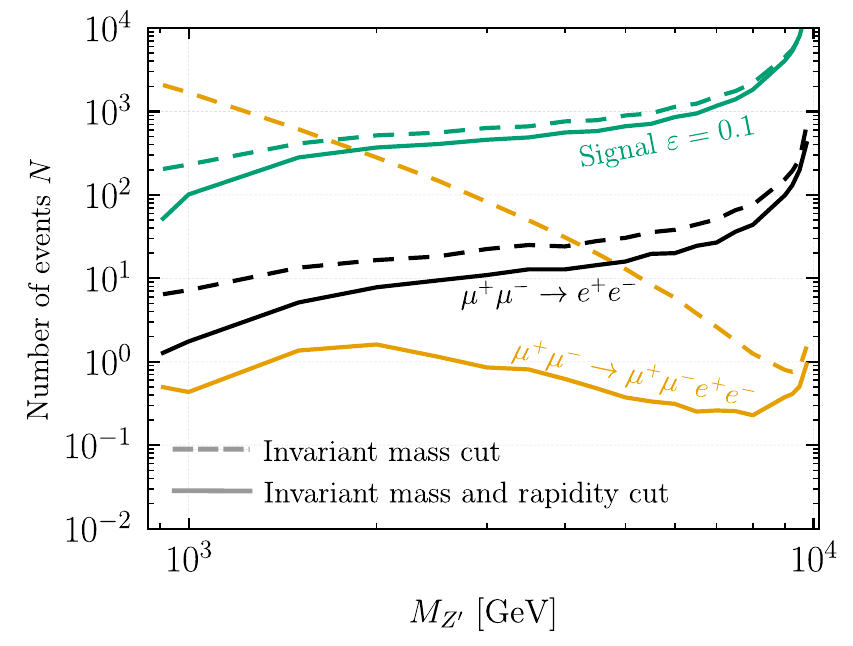}
    \end{minipage}
    \hfill
    \begin{minipage}[t]{0.48\linewidth}
        \centering
        \includegraphics[width=\linewidth]{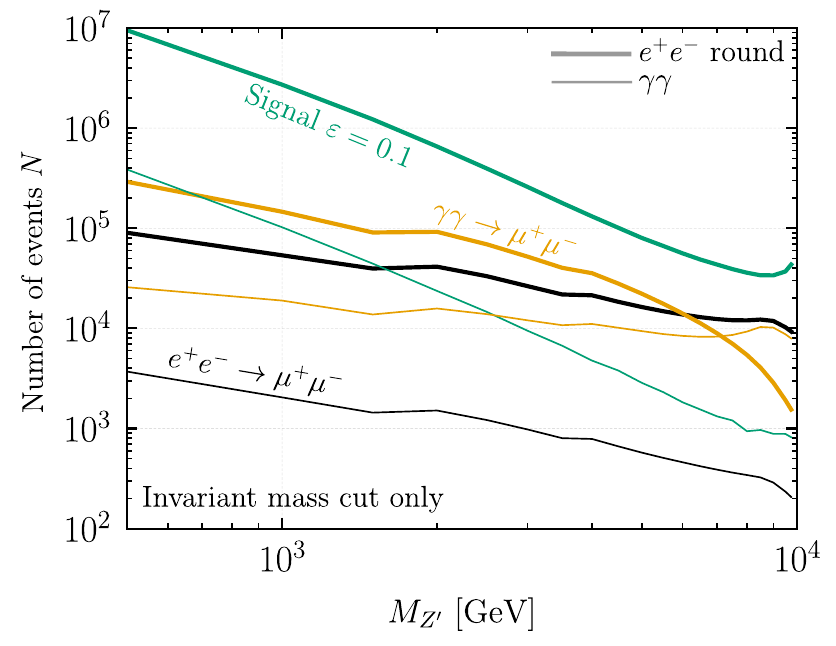}
    \end{minipage}
    \caption{Expected signal (\textcolor{cbgreen}{\textbf{green}}) and 
    background event rates as a function of $M_{Z'}$ for 
    $\varepsilon = 0.1$ and $L_{\text{geom}}=\SI{10}{\per\atto\barn}$. 
    \textbf{(Left)} Inclusive search at a $\SI{10}{\TeV}$ MuC 
    ($\mu^+\mu^- \to e^+e^- + X$). The irreducible background is 
    shown in \textbf{black} and the VBF background in 
    \textcolor{cborange}{\textbf{orange}}; dashed lines include only 
    the invariant mass cut, solid lines add the rapidity cut. 
    \textbf{(Right)} Inclusive search at a WFC for round $e^+e^-$ 
    (thick-solid) and $\gamma\gamma$ (thin-solid) beam configurations. 
    Backgrounds are $e^+e^- \to \mu^+\mu^-$ (\textbf{black}) and 
    $\gamma\gamma \to \mu^+\mu^-$ (\textcolor{cborange}{\textbf{orange}}). 
    All cuts are listed in \cref{tab:cuts}.}
    \label{fig:event_rates}
\end{figure*}

\emph{Irreducible Background:}
The SM process $\ell^+\ell^- \rightarrow \gamma^*/Z^*\rightarrow \ell^{\prime +}\ell^{\prime -}$ constitutes an irreducible background. Integrating over the di-lepton invariant mass window in Eq.~\eqref{eq:dilepton_mass_cut}, the cross section is
\begin{align}
    \sigma^{(\Delta)}_{\rm irr}(M_{Z'}) \approx \frac{4.75\, \alpha^2}{s}2\Delta \frac{d\mathcal{L}_{\ell \ell}}{d\tau}\Big|_{\tau_{Z'}}\,.
\end{align}
The factor 4.75 accounts for both $\gamma^*/Z^*$ diagrams.\\

\emph{Photon Fusion Background:}
The differential cross section for $\gamma\gamma \rightarrow \ell^{\prime +}\ell^{\prime -}$ after the cut on $m_{\ell'\ell'}$ is
\begin{align}\label{eq:sigma_gg_Delta}
    \frac{d\sigma^{(\Delta)}_{\gamma\gamma}(M_{Z'})}{d\cos\theta} \approx \frac{2\pi\alpha^2}{s}2\Delta \frac{d\mathcal{L}_{\gamma \gamma}}{d\tau}\Big|_{\tau_{Z'}}\left( \frac{1+\cos^2\theta}{1-\cos^2\theta} \right)\,,
\end{align}
where $\theta$ is the angle of final state leptons in the CM frame. 
This diverges in the forward limit, but is regularized by pseudorapidity cuts which we now consider.\\

\emph{Acceptance Efficiencies:} The dominant acceptance cut is on pseudorapidity, $|\eta_{\ell'}| < \eta_{\rm max}$. 
Angular distributions for the signal and irreducible background are proportional to $(3/8)(1 + \cos^2 \theta + A_{\rm FB}\cos \theta)$, where $A_{\rm FB}$ is the forward-backward asymmetry; the linear term vanishes upon integration over the symmetric detector volume. 
Collisions with rapidity $y$ in the lab frame have acceptance efficiency
\begin{align} 
    \epsilon_{\ell^+ \ell^-}(y) &= 
        \max\left[0,\,\dfrac{3}{4} \xi(y) + \dfrac{1}{4} \xi(y)^3\right]\,,\\
        \xi(y) &= \tanh(\eta_{\rm max} - |y|)\,,
\end{align}
where the subscript $\ell^+ \ell^-$ indicates this applies to signal and irreducible background with $\ell^+ \ell^-$ in the initial state.

The photon fusion background angular dependence is given by the final term in parentheses in Eq.~\eqref{eq:sigma_gg_Delta}. Integrating this factor over the acceptance range gives\footnote{This is technically a phase-space factor rather than an efficiency, but we use the term efficiency regardless.}
\begin{align}
    \epsilon_{\gamma\gamma}(y) &= \max\left[0,\,4(\eta_{\rm max} - |y|) - 2\xi(y)\right]\,.
\end{align}

We are interested in these efficiencies integrated over rapidities consistent with selection cuts~\cref{eq:dilepton_rapidity_cut}. 
We define the conditional rapidity distribution for fixed $\tau$ as
\begin{align}
    p_{ij}(y|\tau) = \left(\frac{d\mathcal{L}_{ij}}{d\tau\, dy}\bigg/ \frac{d\mathcal{L}_{ij}}{d\tau}\right)\,.
\end{align}
The rapidity-averaged acceptance is
\begin{align}
    \bar{\epsilon}^{(\delta)}_{ij}(M_{Z'}) = 2\int_{y_{Z'} - \delta}^{y_{Z'}} dy\, p_{ij}(y|\tau_{Z'})\, \epsilon_{ij}(y)\,.
\end{align}

At a MuC, the sharply peaked rapidity distribution allows a tight rapidity cut to suppress the photon-fusion background. 
While at a WFC, the broad rapidity distribution prevents this cut from significantly improving sensitivity.\\

\emph{Fiducial Cross Sections:} The fiducial cross sections (neglecting $p_T$ and $\Delta R$ cuts) for processes of interest are
\begin{align}
    \sigma^{\rm fid}_{Z'}(M_{Z'}) &\approx \frac{12\pi^2\, \mu_{Z'}}{s} \left(\frac{d\mathcal{L}_{\ell \ell}}{d\tau}\Big|_{\tau_{Z'}}\right) \bar{\epsilon}^{(\delta)}_{\ell^+\ell^-}(M_{Z'})\,,\\[1ex]
    \sigma^{\rm fid}_{\rm irr}(M_{Z'}) &\approx \frac{4.75\, \alpha^2}{s}\left(2\Delta \frac{d\mathcal{L}_{\ell \ell}}{d\tau}\Big|_{\tau_{Z'}}\right) \bar{\epsilon}^{(\delta)}_{\ell^+\ell^-}(M_{Z'})\,,\\[1ex]
    \sigma^{\rm fid}_{\gamma\gamma}(M_{Z'}) &\approx \frac{2\pi\alpha^2}{s}\left(2\Delta \frac{d\mathcal{L}_{\gamma \gamma}}{d\tau}\Big|_{\tau_{Z'}}\right) \bar{\epsilon}^{(\delta)}_{\gamma\gamma}(M_{Z'})\,.
\end{align}

\emph{Sensitivity:} For masses where the irreducible background dominates (see numerical results in \cref{fig:event_rates}), we obtain the expected 95\% CL upper limit on the signal strength by requiring $S/\sqrt{B} = 1.645$. 
Using $\sqrt{s} = \SI{10}{\TeV}$, $L_{\rm geom} = \SI{10}{\per\atto\barn}$, and setting $\alpha = 1/128$, gives
\begin{align} \label{eq:Inclusive_Search_Sensitivity}
    \mu_{Z'}^{(95\%)}\approx 10^{-8}\,     \left(\frac{d\mathcal{L}_{\ell\ell}}{d\tau}\Big|_{\tau_{Z'}}\right)^{-1/2}\left(   \frac{\Delta}{0.05}\right)^{1/2}\,,
\end{align}
where as an approximation we assumed 100\% signal efficiency. For completeness, the analytic estimate of the exclusive search sensitivity is given in \cref{eq:Exclusive_Search_Sensitivity}, and is similar to \cref{eq:Inclusive_Search_Sensitivity} except with an additional phase-space suppression of roughly $\sqrt{2\eta_{\rm max}/\log(M_{Z'}^2/m_{\ell}^2)}$.

For a WFC, the dominant background for most $Z'$ masses is photon fusion. The sensitivity scales as
\begin{align}
    \mu_{Z'}^{(95\%)}\approx 10^{-8}&\,    \left( \frac{ d\mathcal{L}_{\ell \ell}}{d\tau}\Big|_{\tau_{Z'}}\right)^{-1} \left(   \frac{\Delta}{0.05}\right)^{1/2}\times \notag\\
    &\left(\frac{ d\mathcal{L}_{\gamma \gamma}}{d\tau}\Big|_{\tau_{Z^\prime}}\bar{\epsilon}^{(\delta)}_{\gamma \gamma}(M_{Z^\prime})\right)^{1/2}\,.
\end{align}
\begin{table*}[t!]
\centering
\caption{Summary of acceptance cuts, selection cuts, and 
detector resolutions used in the analysis.}
\label{tab:cuts}
\renewcommand{\arraystretch}{1}
\setlength{\tabcolsep}{16pt}
\begin{tabular}{l c c c}
\toprule
 & \textbf{MuC inclusive} & \textbf{MuC exclusive} & \textbf{WFC} \\
 & $\mu^+\mu^- \to e^+e^- + X$ & $\mu^+\mu^- \to e^+e^-\gamma$ & $e^\pm e^- /\gamma \gamma \to \mu^+\mu^- + X$ \\
\midrule
\multicolumn{4}{l}{\textbf{\emph{Acceptance Cuts}}} \\[2pt]
\quad $p_T^{\min}$  & $\SI{25}{\GeV}$ & $e^\pm:\,\SI{25}{\GeV},\, \gamma:\,\SI{25}{\GeV}$ & $\SI{25}{\GeV}$ \\
\quad $\eta^{\max}$ & 2.5 & 2.5 & 2.5 \\
\quad $\Delta R^{\min}$ & 0.1 & 0.1 & 0.1 \\[4pt]
\midrule
{\textbf{\emph{Resolution}}} & $\sigma(E_e)/E_e = 1\%$ & $\sigma(E_e)/E_e = 1\%$ & $\sigma(1/p_{T\mu}) = \SI{2}{\% \per\TeV}$  \\[4pt]
\midrule
\multicolumn{3}{l}{\textbf{\emph{Di-lepton resolution}}} \\[2pt]
\quad $\sigma_{m_{\ell\ell}}/m_{\ell\ell}$ & $0.7\%$ & $0.7\%$ & $1\% \times m_{\mu\mu}/\text{TeV}$   \\
\quad $\sigma_y$ & $ 0.5\%$ & $ 0.5\%$ & -- \\[4pt]
\midrule
\multicolumn{4}{l}{\textbf{\emph{Selection Cuts}}} \\[2pt]
\quad $\Delta$ (invariant mass) & 0.01 & 0.01 & $0.015 \,m_{\mu \mu}/\text{TeV}$ \\
\quad $\delta$ (rapidity) & $0.01$ & -- & -- \\
    \bottomrule
\end{tabular}
\end{table*}

\emph{Collider Comparison:}
The above expressions reveal three key differences between a MuC and WFC for resonance searches. First, beamstrahlung-induced luminosity spectra at a WFC greatly exceed parton luminosity functions at a MuC ($d\mathscr{L}_{\ell\ell}/d\tau \gg d\mathcal{L}_{\ell\ell}/d\tau$); this provides a much larger signal rate. Second, the conditional rapidity distribution at a MuC is sharply peaked at $|y| = y_{Z'}$; this causes signal acceptance to drop severely when $y_{Z^\prime} > \eta_\mathrm{max}$ i.e. $M_{Z'} < \sqrt{s}\,e^{-\eta_{\rm max}}$. The broad rapidity distribution at a WFC alleviates this loss. Third, tight rapidity cuts available at a MuC efficiently suppress photon fusion backgrounds. At a WFC, the broad rapidity spectrum reduces the effectiveness of this cut and the photon fusion background dominates. 

\subsection{Numerical Analysis} 
\label{sec:Main_analysis}
We now describe the numerical implementation of the search strategies outlined above. Both the MuC and WFC analyses use \texttt{MadGraph5\_aMC@NLO\_v3.6.3}~\cite{Alwall:2014hca} for event generation. 
For all searches we apply the acceptance and selection cuts shown in \cref{tab:cuts}. \\

\emph{Detector Modeling:}
We model detector effects for MuC and WFC searches based on the MUSIC detector concept~\cite{InternationalMuonCollider:2025sys,Delphes_MUSIC}. 
The electron energy resolution and muon $p_T$ resolution are shown in \cref{tab:cuts}. 
Effects on $m_{\ell \ell}$ and $y_{\ell \ell}$ from angular resolution are negligible. 
 
Selection cut windows ($\Delta$, $\delta$) are motivated using 
\begin{align}
    &\frac{\sigma(m_{ee})}{m_{ee}} \approx \sqrt{\frac{5}{2}}\sigma(y_{ee}) \approx \frac{1}{\sqrt{2}}\frac{\sigma(E_e)}{E_e}\,, \\[6pt]
    &\frac{\sigma(m_{\mu\mu})}{m_{\mu \mu}} \approx \sqrt{\frac{3}{10}}\sigma(1/p_{T\mu})\, m_{\mu \mu}\,,
\end{align}
where we have averaged over the decay angles in the resonance rest frame and assumed the resonance is produced in the forward direction for the muon resolution. \\

\emph{Muon Collider:} We perform two searches at the MuC: an inclusive search for di-electron resonances ($\mu^+\mu^- \to e^+e^- + X$) and an exclusive search requiring a tagged photon ($\mu^+\mu^- \to e^+e^-\gamma$).

For the inclusive search, backgrounds include the irreducible process ($\mu^+\mu^- \to \gamma^*/Z^* \to e^+e^-$), neutral-current VBF ($\mu^+\mu^- \to \mu^+\mu^- e^+e^-$, dominated by $\gamma\gamma \to e^+e^-$), and charged-current VBF ($\mu^+\mu^- \to \nu_\ell\bar{\nu}_\ell e^+e^-$, dominated by $W^+W^- \to e^+e^-$). 
The rapidity cut suppresses VBF while retaining signal, as shown in \cref{fig:event_rates}. The exclusive search is dominated by the irreducible background ($\mu^+\mu^-\to e^+e^-\gamma$).

The inclusive search is most sensitive at high masses due to the kinematic enhancement from forward ISR photons. 
At low masses, sensitivity degrades as boosted di-leptons fall outside the detector acceptance and VBF backgrounds grow. 
The exclusive search recovers low-mass sensitivity at the cost of reduced signal rate.\footnote{In the exclusive search one could alternatively cut on the recoil mass $M_{\mathrm{recoil}}=\sqrt{s-2\sqrt{s}E_\gamma}$, which satisfies $M_{\mathrm{recoil}}\approx M_{Z'}$ in the absence of detector effects, see Ref.~\cite{Cheung:2025uaz}. The recoil mass resolution surpasses the di-electron invariant mass resolution for $M_{Z'} \gtrsim \SI{6}{\TeV}$; however, even in this regime the inclusive search remains more sensitive. We therefore use the invariant mass cut throughout.}

For the inclusive search, signal and irreducible background events are generated using the \texttt{isronlyll}~\cite{isronlyll} package which captures the kinematics of forward ISR. 
For $M_{Z'} \lesssim \SI{3}{\TeV}$, the luminosity provided by \texttt{isronlyll} deviates appreciably from the full LePDF~\cite{Garosi:2023bvq} result which solves the DGLAP equations at the double-log order (see \cref{fig:Leff}). 
We therefore re-weight the generated event yields by the ratio of the LePDF to \texttt{isronlyll} luminosities, integrated over the rapidity selection window:
\begin{align}
    N(M_{Z'})=N_{\mathrm{MG}}(M_{Z'})\, 
    \frac{\int_{y_{Z'}-\delta}^{y_{Z'}}dy 
    \frac{d^2\mathcal{L}_\text{LePDF}}{dM\,dy}(M_{Z'})}
    {\int_{y_{Z'}-\delta}^{y_{Z'}}dy
    \frac{d^2\mathcal{L}_{\texttt{isronlyll}}}{dM\,dy}(M_{Z'})}
\end{align}
All effective luminosity curves are detailed in \cref{app:analytic-approximation-details}. The resulting event rates after selection cuts are shown in \cref{fig:event_rates} and \cref{tab:MuC_events}. \\

\begin{figure*}[t!]
    \centering
    \includegraphics[width=0.95\linewidth]{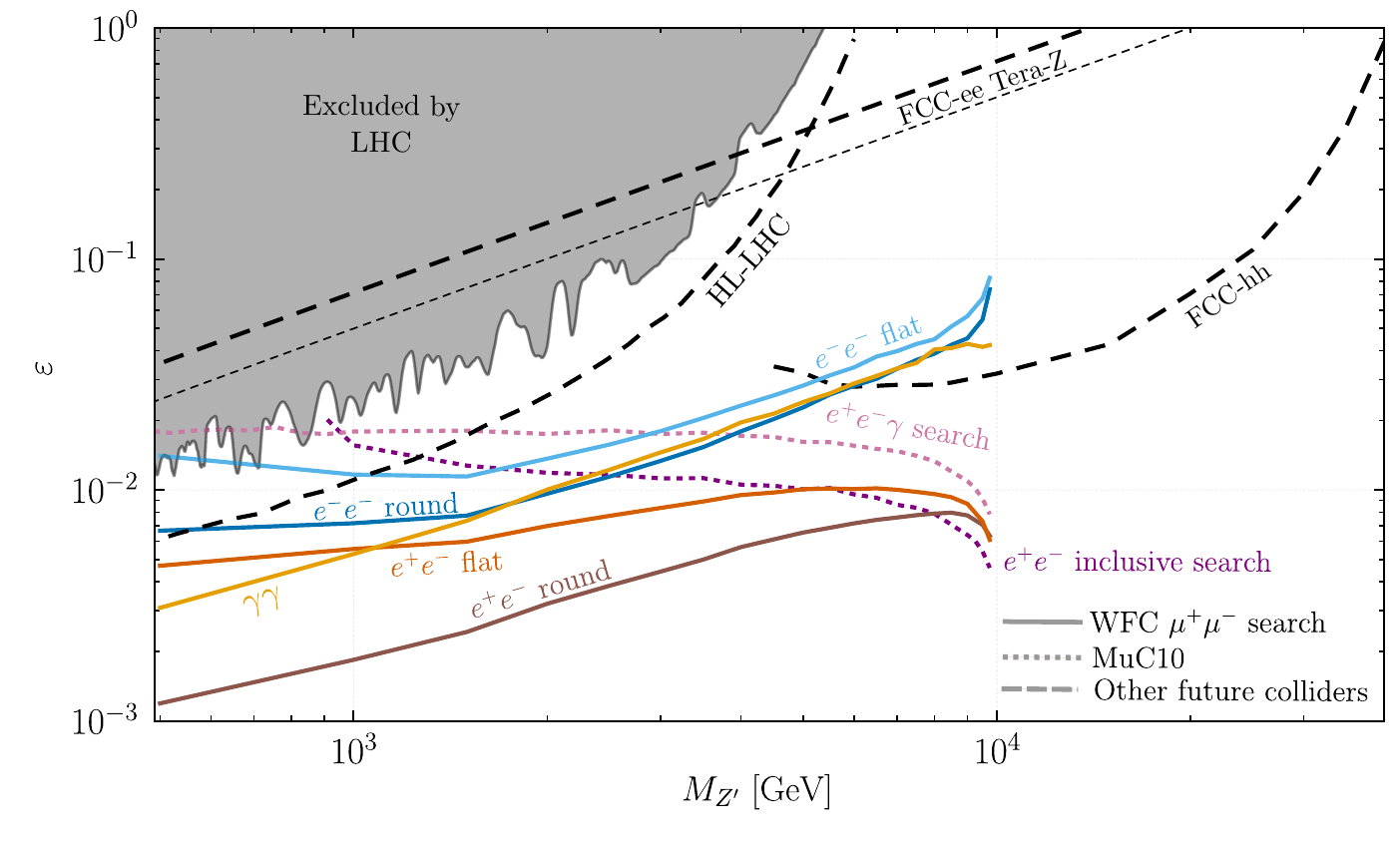}
    \caption{Projected 95\% CL exclusion limits in the $(M_{Z'},\varepsilon)$ plane for a kinetically mixed $Z'$. 
    The grey shaded region shows existing LHC constraints ($\sqrt{s}=\SI{13}{\TeV}$, $L_{\rm int}=\SI{139}{\per\femto\barn}$) \cite{LHC}. Dotted curves show a $\SI{10}{\TeV}$ MuC with $\mathcal{L}=\SI{10}{\per\atto\barn}$; solid curves show WFC configurations with ${L}_{\text{geom}}=\SI{10}{\per\atto\barn}$. Dashed black curves show projections for the HL-LHC ($L_{\rm int}=\SI{3}{\per\atto\barn}$), FCC-hh ($\sqrt{s}=\SI{100}{\TeV}$, $L_{\rm int}=\SI{30}{\per\atto\barn}$) \cite{FCC_hh}, and electroweak precision tests at FCC-ee \cite{fccee,T_from_ZP}.}
    \label{fig:Results_plot}
\end{figure*}

\emph{Wakefield Collider:} At a WFC we search for di-muon resonances mediated by $e^+e^- \to Z' \to \mu^+\mu^-$, evaluated using the effective $e^+e^-$ luminosity
generated through beamstrahlung. 
For the primary $e^+e^-$ configuration this luminosity is carried mainly by primary beam leptons after beamstrahlung, whereas for the $e^-e^-$ and $\gamma\gamma$ configurations the dominant contribution comes from secondary $e^+e^-$ pairs produced in the interaction region. 
In particular, the direct production channel $\gamma\gamma \to Z^\prime + X$ is suppressed, see \cref{fig:All_PWFA_events}. 
Unlike the MuC, only an inclusive search is performed: the broad rapidity distribution keeps signal events sufficiently central that the exclusive search is not needed to recover low-mass sensitivity.

Backgrounds are analogous to the MuC inclusive search with lepton flavors interchanged.
The irreducible background ($e^+e^- \to \gamma^*/Z^* \to \mu^+\mu^-$) is treated identically, but evaluated using the corresponding effective $e^+e^-$ luminosity
spectrum. 
Photon fusion ($\gamma\gamma \to \mu^+\mu^-$), sourced by the on-shell photon luminosity in the interaction region, contributes significantly for all collider configurations. 
Unlike at the MuC, the broad rapidity spectrum (\cref{fig:Rapidity_Comparison_Combined}) prevents a rapidity cut from effectively suppressing this background without excessive signal loss; we therefore do not apply
the cut on $y_{\mu^+\mu^-}$.

In contrast to the muon collider search, where events are generated at a single center-of-mass energy $\sqrt{s}$, WFC analyses must account for the broad spectrum of collision energies produced by beamstrahlung. Following the approach in Ref.~\cite{PWFA}, we approximate this by generating events with \texttt{MadGraph5} on a uniform grid in $\sqrt{\hat{s}}$ from $\SI{0.5}{\TeV}$ to $\SI{10}{\TeV}$ in steps of $\SI{50}{\GeV}$, and reweighting each sample by the beamstrahlung luminosity integrated over the corresponding energy bin $\mathscr{L}^{\rm bin}(\sqrt{\hat{s}})$.\footnote{The $\sqrt{\hat{s}} = \SI{10}{\TeV}$ bin also includes the $\delta$-function contribution at the nominal collision energy.} 
The total background yield for a given $M_{Z'}$ hypothesis is then
\begin{align}\label{eq:N_bkg_WFC}
    N_{\rm bkg}(M_{Z'}) = L_{\rm geom} 
      \sum_{\text{bins}}
    \mathscr{L}^{\rm bin}(\sqrt{\hat{s}})\,\sigma(\sqrt{\hat{s}})
    \,\epsilon(M_{Z'})\,,
\end{align}
where the sum runs in $\SI{50}{\GeV}$ steps, $\sigma$ is the partonic cross section at $\sqrt{\hat{s}}$ and $\epsilon(M_{Z'})$ is the efficiency of all acceptance and selection cuts, including the rapidity distribution, evaluated at that energy. Background cross sections vary slowly with $\sqrt{\hat{s}}$, so the dominant contribution comes from bins near $\sqrt{\hat{s}} \approx M_{Z'}$, with neighboring bins entering through the finite invariant mass window. For signal, events are generated at $\sqrt{\hat{s}} = M_{Z'}$ and the cross section is evaluated using the narrow-width approximation. The resulting event rates are shown in \cref{fig:event_rates} and \cref{tab:PWFA_events}.
\subsection{Results}
\label{subsec:Results}
\begin{figure}[t!]
    \centering
    \includegraphics[width=\linewidth]{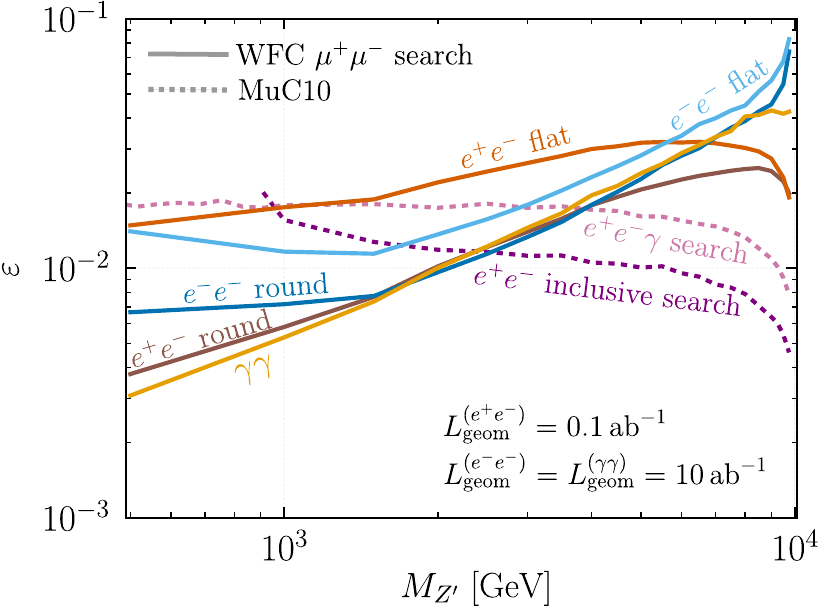}
    \caption{Same as \cref{fig:Results_plot}, but with geometric luminosities adjusted to reflect the greater difficulty of producing positron beams at a WFC. The $e^+e^-$ configurations assume $L_{\rm geom} = \SI{0.1}{\per\atto\barn}$, while the $e^-e^-$ and $\gamma\gamma$ configurations retain $L_{\rm geom} = \SI{10}{\per\atto\barn}$.}
    \label{fig:Reduced_Results_plot}
\end{figure}

Projected 95\% CL exclusion limits for a kinetically mixed $Z'$ are shown in \cref{fig:Results_plot}. We summarize the main findings below. To underscore the difficulty of positron acceleration we also show in \cref{fig:Reduced_Results_plot} a version with the $e^+e^-$ luminosity for both round and flat beams reduced to $L_\text{geom} = \SI{0.1}{\per\atto\barn}$. 

\emph{Muon Collider:} At the MuC, the inclusive search provides the strongest sensitivity for $M_{Z'} \gtrsim \SI{1}{\TeV}$, benefiting from the enhancement of ultra-forward ISR. 
The exclusive search extends coverage to lower masses where the inclusive search loses acceptance. Together, a $\SI{10}{\TeV}$ MuC with $L_{\rm geom} = 10~\text{ab}^{-1}$ can probe $\varepsilon \gtrsim 10^{-2}$ across $\SI{1}{\TeV} \lesssim M_{Z'} \lesssim \SI{10}{\TeV}$, exceeding HL-LHC projections throughout this range.
Our results are stronger than those of Ref.~\cite{Airen:2024iiy} in the overlapping mass regime; that analysis considered same-flavor leptons in the initial and final states, leading to larger backgrounds. 
In addition, our limits are slightly stronger than those of Ref.~\cite{Cheung:2025uaz}, which instead performs an exclusive search using the recoil mass approach. 
Finally, we find good agreement between our results and Ref.~\cite{Dasgupta:2023zrh} at comparable collider energies.

\emph{Wakefield Collider:} WFC configurations exhibit significantly improved sensitivity for lower masses, reflecting the large beamstrahlung-induced luminosity. The $e^+e^-$ round configuration provides the strongest limits of $\varepsilon \gtrsim 10^{-3}$ for $M_{Z'} \sim \SI{500}{\GeV}$, an order of magnitude better than a MuC at the same mass. The advantage diminishes as $M_{Z'} \to \sqrt{s}$, where beamstrahlung provides less enhancement and the di-muon resolution becomes worse than the di-electron resolution. In \cref{fig:Reduced_Results_plot} we see that the $\gamma\gamma$ and $e^-e^-$ variants perform similarly in the case that the positron luminosity is two orders of magnitude smaller than what is achievable for electrons. We note that the $\gamma\gamma$ collider performs comparably to or better than the $e^-e^-$ collider across the full mass range.
 
\emph{Other Colliders:} For comparison, we include existing LHC constraints \cite{LHC} and projections for the HL-LHC and FCC-hh \cite{FCC_hh}. For HL-LHC we take the expected LHC limit \cite{LHC} and rescale it from $L_{\mathrm{int}}=\SI{139}{\per\femto\barn}$ to $L_{\mathrm{int}}=\SI{3}{\per\atto\barn}$. The FCC-hh leverages its high energy to extend coverage beyond $M_{Z'} \gtrsim \SI{10}{\TeV}$, but does not match the WFC sensitivity for $M_{Z'} \lesssim \SI{5}{\TeV}$. This reflects the large beamstrahlung enhancement of resonance production at a WFC, which grows by several orders of magnitude toward lower masses (cf.\ \cref{fig:Rratio}), combined with the comparatively modest backgrounds afforded by a clean leptonic initial state. We also show projections from electroweak precision measurements at FCC-ee, which provide complementary sensitivity to virtual $Z'$ effects through the oblique parameter $\hat{T}$ \cite{T_from_ZP}. The reach depends on progress in reducing theoretical uncertainties; we show both conservative and optimistic scenarios \cite{fccee}.

\section{Conclusion}
\label{sec:Conclusion}
Beamstrahlung at a wakefield collider, often viewed as a drawback, provides a powerful advantage for resonance searches by redistributing luminosity across a broad energy spectrum. For the benchmark of a kinetically mixed $Z^\prime$, a $\SI{10}{\TeV}$ WFC can improve sensitivity to $\varepsilon$ by an order of magnitude compared to a MuC with the same nominal energy and geometric luminosity. The framework developed here can be applied to other resonances beyond the kinetically mixed $Z'$. The narrow-width approximation makes it straightforward to recast these projections for any narrow resonance coupling to leptons or photons.

More broadly, the beamstrahlung enhancement ratio introduced in \cref{subsec:Luminosity Spectra} shows that all processes with cross sections weighted toward low energies benefit significantly from beamstrahlung, with resonance production receiving the largest enhancement. This pattern is directly tied to the shape of the luminosity spectra and is largely independent of the specific model considered.

The $e^+e^-$ configurations provide the strongest sensitivity, underscoring the value of positron beams. An $e^-e^-$ or $\gamma\gamma$ WFC would require significantly more integrated luminosity to match the reach of $e^+e^-$ machines. However, large beamstrahlung at a WFC partially compensates for the absence of a primary positron beam as many positrons are generated by beam-beam interactions and give both $e^-e^-$ and $\gamma\gamma$ colliders better sensitivity than one might naively expect. Production of on-shell photons provides substantial $\gamma\gamma$ luminosity at all collider configurations. This could be exploited for physics targets beyond di-lepton resonances studied here, such as photophilic axion-like particles which we leave for future work. 

We note that this advantage is specific to bump-hunt searches where the resonance mass is reconstructed from final-state 
kinematics. 
Many important measurements at lepton colliders rely on precise knowledge of the collision energy on an event-by-event basis---including threshold scans, recoil-mass measurements, and kinematic endpoint analyses---and these are degraded by the large beamstrahlung energy spread. 
Such techniques remain a unique strength of Higgs factories and muon colliders, where the beam energy spread is small and well characterized. 
The physics programs of wakefield and muon colliders are therefore largely complementary: wakefield colliders excel at scanning broad mass ranges for weakly coupled resonances, while muon colliders offer unmatched precision at a chosen center-of-mass energy.

As the particle physics community evaluates options for the next energy-frontier facility, it is important to understand the full physics potential of each proposal. Wakefield colliders offer a qualitatively different collision environment from both hadron and muon colliders, and the present work demonstrates that this environment can be a significant asset for discovery. 

\addtocontents{toc}{\protect\hidetocsectionstrue}
\begin{acknowledgments}
We are grateful to Simon Knapen and Michael Peskin for valuable feedback on the manuscript and to David Marzocca for helpful discussions.
This work builds on an ongoing collaboration with colleagues in the LBNL Accelerator Technology \& Applied Physics Division and the LBNL ATLAS group. We are especially grateful to Stepan Bulanov, Arianna Formenti, Remi Lehe, Jens Osterhoff, Carl Schroeder and Jean-Luc Vay for sharing their preliminary accelerator parameters and beam–beam simulation results, and to Kehang Bai, Simone Pagan Griso and Angira Rastogi for guidance on detector-related assumptions. We thank all of the above for numerous discussions that sharpened our understanding and for their comments on the manuscript.
\end{acknowledgments}

\addtocontents{toc}{\protect\tocappendixtrue}
\addtocontents{toc}{\protect\hidetocsectionsfalse}
\addtocontents{toc}{\protect\hidetocsubsectionstrue}

\addtocontents{toc}{\protect\hidetocsectionstrue}
\bibliographystyle{JHEP}
\bibliography{mybibliography}

\appendix

\clearpage 

\onecolumngrid

\setlength{\parskip}{6pt}

\section{Analytic Approximation Details}
\label{app:analytic-approximation-details}

This appendix derives approximate analytic expressions for signal and background rates in resonance searches, clarifying the parametric dependence and features of projections. We employ a framework for calculations of resonance production via radiative return which easily extends to resonance production via beamstrahlung; this requires only the substitution of parton luminosity functions with the appropriate beamstrahlung spectra.

\subsection{Parton distribution functions}
We treat incoming particles as a collection of collinear partons. The \emph{parton distribution functions} $f_{a/i}(x,Q^2)$ describe the distribution of parton $a$ coming from parent particle $i$ with energy fraction $x$ evaluated at a factorization scale $Q^2$. The \emph{parton luminosity function} for collisions of partons $a$ and $b$ from parent particles $i$ and $j$ is
\begin{align} \label{eq:dLdtaudy}
    \frac{d\mathcal{L}_{ab/ij}}{d\tau d y } =  f_{a/i}\left(\sqrt{\tau}e^y,Q^2\right) f_{b/j}\left(\sqrt{\tau}e^{-y},Q^2\right) + (a\leftrightarrow b~{\rm if}~a \neq b).
\end{align}
Parent particle indices are suppressed when unambiguous. Here, $ \tau = x_1 x_2 =  \hat{s}/s$ where $\sqrt{\hat{s}}$ is the parton center-of-mass energy. Additionally, $y=\frac{1}{2}\log\left(x_1/x_2\right)$ is the parton rapidity with kinematic limits $|y| \leq y_{\rm max}(\tau) = \log(1/\sqrt{\tau})$. 

Since resonance searches select a specific parton center-of-mass energy $\tau_X = M_X^2/s$, it is convenient to define the \emph{conditional rapidity distribution} for fixed $\tau$ as
\begin{align}\label{eq:p_of_y}
    p_{ab}(y|\tau) = \left(\frac{d\mathcal{L}_{ab}}{d\tau dy} \bigg/ \frac{d\mathcal{L}_{ab}}{d\tau}\right).
\end{align}

When initial-state leptons radiate collinear photons prior to collision, the effective center-of-mass energy is reduced. At first order (single photon emission), the lepton PDF is:
\begin{align} \label{eq:I_order_Lumi}
    f_{\ell/\ell}(x,Q^2) &\approx  \delta(1-x) +  \frac{ \alpha}{2\pi} \log \left( \frac{Q^2}{m_\ell^2} \right) \left[\frac{1+x^2}{(1-x)_+}  + \frac{3}{2}\delta(1-x)\right],
\end{align}
where the plus distribution is defined by $\displaystyle\int_0^1 dx\, \dfrac{f(x)}{(1-x)_+} = \displaystyle\int_0^1 dx\, \dfrac{f(x)-f(1)}{1-x}$. The parton luminosity function is
\begin{align}\label{eq:dL_ell_ell_dtau_dy}
    \frac{d\mathcal{L}_{\ell \ell/\ell \ell}}{d\tau dy} \approx  \frac{ \alpha}{\pi} \log \left( \frac{Q^2}{m_\ell^2} \right) \frac{1+\tau^2}{(1-\tau)}\, p_{\ell \ell}(y|\tau) + \mathcal{O}(\alpha^2).
\end{align}
The conditional rapidity distribution is approximated by
\begin{align} \label{eq:p_ell_ell_of_y}
    p_{\ell \ell/\ell \ell}(y|\tau) = \frac{1}{2}\delta\big(|y| - y_{\rm max}(\tau)\big).
\end{align}
This distribution peaks at the rapidity limit, corresponding to collisions involving one lepton with maximal energy.

\begin{figure*}[ht!]
    \centering

    \subfloat[Muon beams\label{fig:Luminosity_muons}]{
        \includegraphics[width=0.48\linewidth]{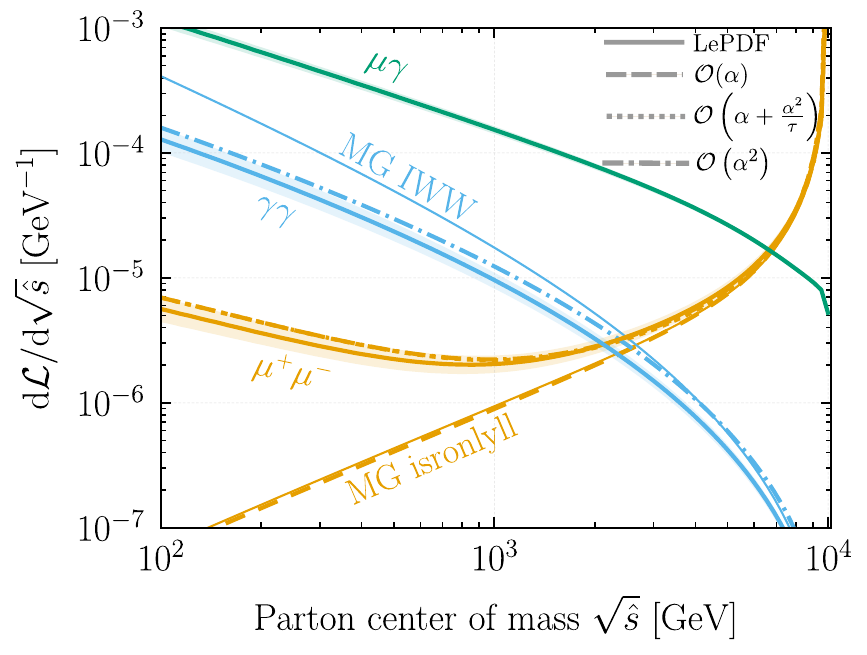}
    }
    \hfill
    \subfloat[Electron beams\label{fig:Luminosity_electrons}]{
        \includegraphics[width=0.48\linewidth]{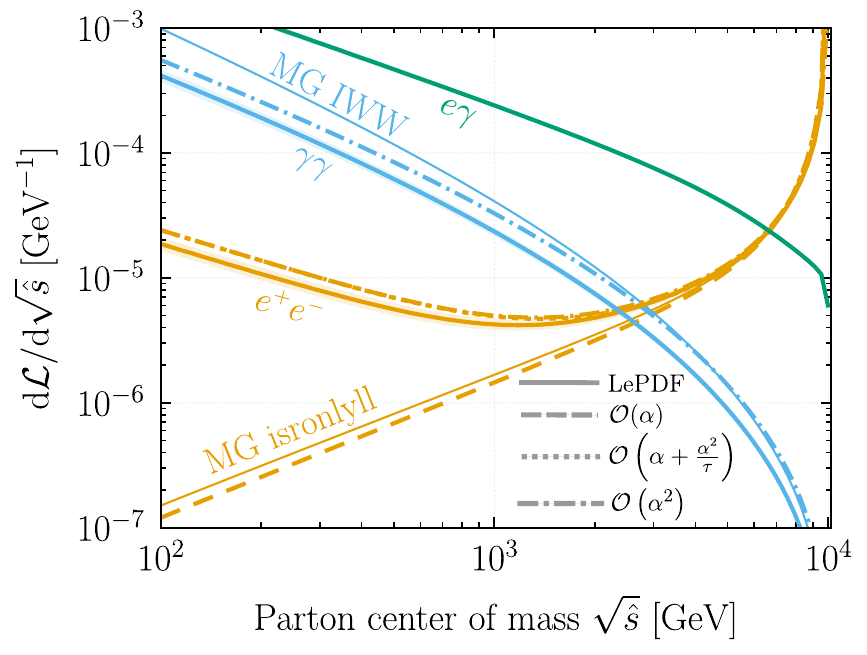}
    }
    \caption{\textbf{(Left)} Parton luminosities at a $\SI{10}{\TeV}$ MuC as a function of $\sqrt{\hat{s}}$. Solid curves with shaded error bands show the full electroweak PDF set (LePDF~\cite{Garosi:2023bvq}) for $\mu^+\mu^-$ (\textcolor{cborange}{\textbf{orange}}), $\gamma\gamma$ (\textcolor{cblblue}{\textbf{light-blue}}), and $\mu\gamma$ (\textcolor{cbgreen}{\textbf{green}}). Dashed, dotted, and thin curves show the analytic approximations described in the text. \textbf{(Right)} The same for electron beams; the smaller lepton mass yields a larger luminosity. The shaded bands indicate the uncertainty from varying the factorization scale between $\sqrt{\hat{s}}/4$ and $\sqrt{\hat{s}}$. 
    }
    \label{fig:Leff}
\end{figure*}

We also consider the distribution of collinear photons radiated from initial-state leptons:
\begin{align} \label{eq:I_order_Lumi_gamma}
    f_{\gamma /\ell}(x,Q^2) &=\frac{ \alpha}{2\pi} \log \left( \frac{Q^2}{m_\ell^2} \right) \frac{1+(1-x)^2}{x} + \mathcal{O}(\alpha^2).
\end{align}
The leading-order luminosity spectrum for photon collisions is:
\begin{align}
    \frac{d\mathcal{L}_{\gamma \gamma/\ell \ell}}{d\tau dy} &= \left[\frac{ \alpha}{2\pi} \log \left( \frac{Q^2}{m_\ell^2} \right)\right]^2 \left[\frac{(2+\tau )^2 - 4(2+\tau ) \sqrt{\tau } \cosh (y)+4 \tau  \cosh (2 y)}{\tau }\right] + \mathcal{O}(\alpha^3).
\end{align}
The $\tau$ and $y$ dependence of this function does not factorize. Integrating over rapidity gives:
\begin{align} \label{eq:photon_lumi}
    \frac{d\mathcal{L}_{\gamma \gamma/\ell \ell}}{d\tau} &=\left[\frac{ \alpha}{2\pi} \log \left( \frac{Q^2}{m_\ell^2} \right)\right]^2 \left[\frac{2 \left(\tau -1\right)\left(\tau +3\right)-(\tau +2)^2 \log (\tau )}{\tau }\right] + \mathcal{O}(\alpha^3).
\end{align}
The photon parton luminosity peaks at low energies and has a more central conditional rapidity distribution:
\begin{align}
    p_{\gamma \gamma/\ell \ell}(y|\tau) = \frac{(2+\tau )^2 - 4(2+\tau ) \sqrt{\tau } \cosh (y)+4 \tau  \cosh (2 y)}{ 2 \left(\tau -1\right)\left(\tau +3\right)-(\tau +2)^2 \log (\tau ) }.
\end{align}

The leading order expressions above capture the qualitative features of the luminosity functions but can deviate significantly from the full result at low masses. 
\Cref{fig:Leff} compares the parton luminosities at several levels of approximation against the resummed LePDF~\cite{Garosi:2023bvq} benchmark (solid curves with shaded error bands). 
See also Refs.~\cite{Han:2020uid,Han:2021kes} for a recent similar PDF set. 
For the $\ell^+\ell^-$ channel, the dashed line shows the $\mathcal{O}(\alpha)$ result from \cref{eq:I_order_Lumi} and the dotted line includes $\mathcal{O}(\alpha^2)$ corrections; for the $\gamma\gamma$ channel, the dash-dotted line shows the $\mathcal{O}(\alpha^2)$ form of \cref{eq:photon_lumi}. 
The thinnest curves correspond to the default \texttt{MadGraph5} settings (using \texttt{isronlyll}~\cite{isronlyll} for leptons and the IWW approximation~\cite{iww} for photons). 
The $\mathcal{O}(\alpha)$ approximation agrees well with LePDF for $\sqrt{\hat{s}} \gtrsim \SI{3}{\TeV}$ but underestimates the luminosity at lower energies, where resummed higher-order contributions become important. 
These deviations motivate the reweighting procedure applied to the MuC inclusive search, described in \cref{sec:Main_analysis}.


\subsection{\textbf{Resonance Production Cross Sections and Acceptance}}
Consider partons $a$ and $b$ forming resonance $X$ that decays to final state $\mathcal{F}$. Assuming the \emph{narrow-width approximation} ($\Gamma_X\ll M_X$), and defining the spin-averaging factor\footnote{Helicities require careful treatment for polarized beams or heavy vector partons, but this is not required for the current analysis.} $\mathcal{S} = \frac{(2S_X+1)}{(2S_a+1)(2S_b+1)}$, the total cross section is
\begin{align}\label{eq:sigma_X_Truth}
    \sigma_{ab\rightarrow X\rightarrow \mathcal{F}}(s) \approx \frac{16\pi^2\, \mathcal{S}\, \mu_X  }{ s} \frac{d\mathcal{L}_{ab}}{d\tau}\Big|_{\tau = \tau_X},
\end{align}
where we defined the dimensionless signal strength
\begin{align}
    \mu_X \equiv \frac{\Gamma(X\rightarrow a b)\,{\rm BR}(X\rightarrow \mathcal{F})}{M_X}.
\end{align}

A finite detector coverage $|\eta| < \eta_{\mathrm{max}}$ limits the acceptance. For a system with rapidity $y$, both particles in a two-body final state are detected only if the decay angle in the rest frame satisfies $|\cos \theta| < \xi(y) \equiv \tanh(\eta_{\mathrm{max}} - |y|)$. 

The differential cross section with respect to the decay angle depends on the spin. Scalar resonances ($S_X=0$) decay isotropically. Vector resonances ($S_X=1$) decay to fermions with a distribution proportional to $1 + \cos^2\theta + A_{\rm FB} \cos\theta$. Integrating over the allowed range yields the acceptance efficiency:
\begin{align} \label{eq:epsilon_general}
    \epsilon_{X}(y) &= 
    \begin{cases}
        \max\left[0,\,\xi(y)\right] ,& S_X = 0 \quad (\text{Scalar})\\[2ex]
        \max\left[0,\,\dfrac{3}{4} \xi(y) + \dfrac{1}{4} \xi(y)^3\right], & S_X = 1 \quad (\text{Vector})
    \end{cases}
\end{align}
Since the detector volume is symmetric, the term linear in $\cos\theta$ vanishes. The fiducial cross section is
\begin{align}
    \sigma^{{\rm fid}}_{ab\rightarrow X\rightarrow \mathcal{F}}(s) = \sigma_{ab\rightarrow X\rightarrow \mathcal{F}}(s) \int p_{ab}(y|\tau_X)\,\epsilon_X(y)\, dy.
\end{align}
Standard Model background processes of the form $ab\rightarrow \mathcal{F}$ can be written in a similar form as seen below.

\subsection{ Inclusive Search ($\ell^+\ell^- \to X\rightarrow {\ell'}^+{\ell'}^-$)} \label{sec:Inclusive Resonance Production}

A resonance $X$ coupled to leptons can be produced via the emission of an unresolved ISR photon that escapes down the beam pipe. This results in a system boosted with rapidity $y_X \equiv y_{\rm max}(\tau_X) = \log(\sqrt{s}/M_X)$.

\subsubsection*{\textbf{Event Selection}}

We define a cut-and-count analysis based on the following selection and acceptance cuts:
\begin{enumerate}
    \item \textbf{Di-lepton Invariant Mass:} We select a window around the di-lepton invariant mass peak determined by the detector leptonic energy resolution $\Delta \approx \sigma(E_{\ell'})/E_{\ell'}$ (this choice maximizes $S/\sqrt{B}$ for flat backgrounds):
    \begin{align}
        |m_{\ell' \ell'} - M_X| < \Delta \, M_X.
    \end{align}
    \item \textbf{Di-lepton Rapidity:} As the signal peaks at $|y_{\ell'\ell'}| = y_X$, we apply a cut at the kinematic edge:
    \begin{align}
        \Big||y_{\ell' \ell'}| - y_X\Big| < \delta.
    \end{align}
    \item \textbf{Pseudorapidity Acceptance:} We require the final state leptons to satisfy:
    \begin{align}
        |\eta_{\ell'}| < \eta_{\rm max}.
    \end{align}
\end{enumerate}
The pseudorapidity cut implies the detector loses sensitivity for $M_X < \sqrt{s} e^{-\eta_{\rm max}}$. Acceptance cuts on $p_T$ and $\Delta R_{\ell'\ell'}$ are not as consequential and are ignored in this approximate analysis.

\subsubsection*{\textbf{Background Rates}}

\begin{figure}
    \centering
    \includegraphics[width=0.6\linewidth]{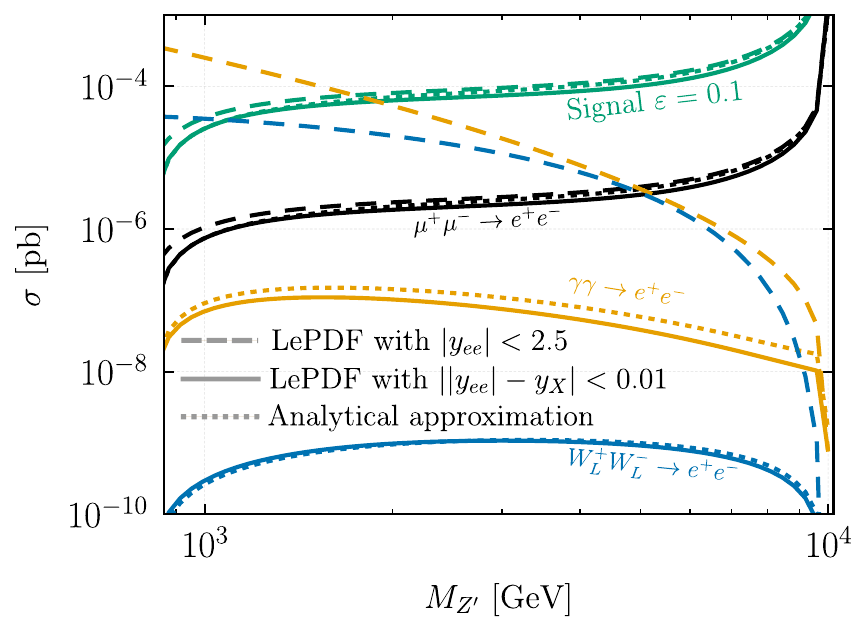}
    \caption{The relevant backgrounds for the inclusive search $\mu^+\mu^-\to e^+e^-$ at a \SI{10}{\TeV} MuC. The irreducible background is represented in \textbf{black}; the signal (\textcolor{cbgreen}{\textbf{green}}) computed at $\varepsilon=0.1$. The reducible backgrounds are neutral current VBF (\textcolor{cborange}{\textbf{orange}}) and charged current VBF (\textcolor{cbblue}{\textbf{blue}}). The dashed lines show the result using LePDF \cite{Garosi:2023bvq} with no selection cuts other than the invariant mass cut $M_X(1-\Delta) < M_{ee} < M_X(1+\Delta)$, with $\Delta=0.01$, and the rapidity cut $|y_{ee}|<2.5$. The solid lines include as well the rapidity cut $||y_{ee}|-y_X|<0.01$ for optimising the signal. The dotted lines show the analytical approximations derived in \cref{sec:Inclusive Resonance Production}.}
    \label{fig:LEPDFvsAnalytical}
\end{figure}

\emph{\textbf{Irreducible Background}}\,: The Standard Model process $\ell^+\ell^- \rightarrow \gamma^*/Z^*\rightarrow \ell^{\prime\, +}\ell^{\prime\,-}$ constitutes an irreducible background. For $\hat{s}\gg M_Z^2$, the differential partonic cross section in terms of the center-of-mass decay angle is proportional to $\left(1 + \cos^2 \theta + A_{\rm FB}\cos \theta \right)$, where $A_{\rm FB}\approx 0.6$. We therefore use the generic vector efficiency $\epsilon_{S=1}(y)$.
Integrating the differential luminosity over the mass window $\Delta$, and assuming the background kinematics generally follow the signal rapidity distribution $p_{\ell \ell}(y|\tau_X) = \frac{1}{2}\delta(|y| - y_X)$, the fiducial cross section is:
\begin{align}
    \sigma^{{\rm fid}}_{\rm irr}(M_X) \approx \frac{4.75 \alpha^2}{s}\left( 2\Delta \frac{d\mathcal{L}_{\ell \ell}}{d\tau}\Big|_{\tau_X}\right) \epsilon_{S=1}\big(y_X\big).
\end{align}
Here, the factor 4.75 accounts for $\gamma^*/Z^*$ interference and couplings.

\emph{\textbf{Vector Boson Fusion (VBF) Background}}\,:
The VBF background is dominated by photon fusion, $\gamma\gamma \rightarrow \ell^{\prime+}\ell^{\prime-}$. Unlike the signal, this process is dominated by $t$-channel exchange. For $\hat{s} \gg m_{\ell'}^2$ the differential cross section is:
\begin{align}
    d\hat{\sigma}_{\gamma\gamma}(\hat{s}) \approx \frac{2\pi\alpha^2}{\hat{s}} \left( \frac{1+\cos^2\theta}{1-\cos^2\theta} \right) d\cos\theta.
\end{align}
The final state leptons are more forward in the parton center-of-mass frame than for the signal. The acceptance factor specific to this process is obtained by integrating the term in parentheses over $|\cos \theta| < \xi(y)$, yielding:
\begin{align}
    \epsilon_{\gamma\gamma}(y) &= \max\left[0,\,4\tanh^{-1}(\xi(y)) - 2\xi(y)\right].
\end{align}
The fiducial cross section is obtained by integrating over the rapidity window, $||y| - y_X| < \delta$. Since the photon luminosity $p_{\gamma\gamma}(y|\tau)$ vanishes beyond the kinematic limit, the integration reduces to single-sided regions at the kinematic endpoints. We approximate the integrand as constant over this range, since the relative error is $\mathcal{O}(\delta \cdot  \tau_X)$, which gives
\begin{align}
    \sigma^{{\rm fid}}_{\gamma\gamma}(M_X) \approx \frac{2\pi \alpha^2}{s}\left( 2\Delta \frac{d\mathcal{L}_{\gamma \gamma}}{d\tau} \Big|_{\tau_X} \right)\,\Big[ 2\delta \, p_{\gamma\gamma}\big(y_X\big) \, \epsilon_{\gamma\gamma}\big(y_X\big) \Big].
\end{align}
Although the $\gamma\gamma$ parton luminosity scales as $\alpha \log(Q^2/m_\ell^2)$ relative to the irreducible background, it also scales as $\tau^{-1}$ for $\tau \ll 1$. Consequently, the photon fusion background tends to dominate for low-mass resonances.

\subsubsection*{\textbf{Inclusive Search Sensitivity}}

\begin{figure}
    \centering
    \includegraphics[width=0.6\linewidth]{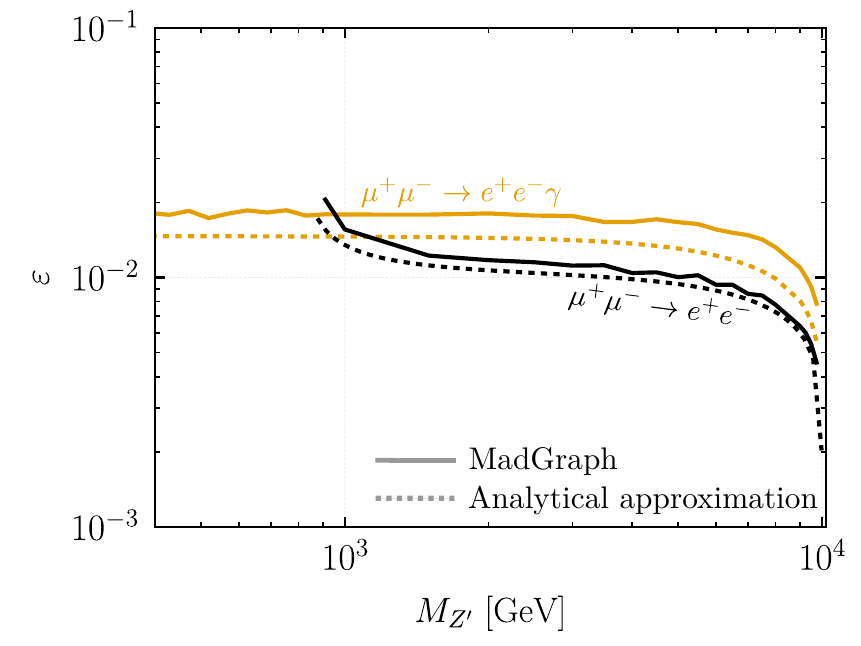}
    \caption{The sensitivity of the two searches we considered at the Muon Collider: \textbf{\textcolor{cborange}{exclusive}} and \textbf{inclusive}. Analytical approximations are also shown here: the orange dotted line refers to \cref{eq:Exclusive_Search_Sensitivity} while the dotted black line uses the full LePDF result \cite{Garosi:2023bvq} (and analytical cross-sections). The latter would have been the same if we had used just the $\mathcal{O}(\alpha^2)$ approximation of LePDF.}
    \label{fig:sensitivity_MuC}
\end{figure}

As shown numerically in the main text, the irreducible background is the dominant background for resonance masses where the inclusive search is optimal for a MuC. Assuming this situation, we estimate the expected 95\% CL upper limit on $\mu_X$ by setting $S/\sqrt{B} = 1.645$ where $S$ ($B$) is the total signal (background) events passing cuts for integrated luminosity $L_{\rm int}$ (i.e. $L_{\rm int}\,\sigma^{{\rm fid}}$). Using $\alpha = 1/128$ for simplicity gives
\begin{align} \label{eq:Inclusive_analytic}
    \mu_X^{(95\%\, \rm UL)}\approx 10^{-8}\, \left(\frac{\Delta}{0.05}\right)^{1/2} \left(\frac{\sqrt{s}}{10~{\rm TeV}}\right) \left(  \frac{  L_{\rm int} }{10~{\rm ab}^{-1}}  \frac{ d\mathcal{L}_{\ell \ell}}{d\tau}\Big|_{\tau_X} \epsilon_{S = 1}(y_X)\right)^{-1/2}.
\end{align}
This result is presented in \cref{fig:sensitivity_MuC} for the kinetically mixed $Z'$ along with results from performing numerical simulations.

\subsection{Exclusive Search ($\ell^+\ell^- \to \gamma (X\rightarrow {\ell'}^+{\ell'}^-$))} \label{sec:Exclusive Resonance Production}

The sharp decline in acceptance for low-mass resonances and large VBF backgrounds motivates searching for a resonance $X$ produced in association with a hard, resolved photon. The hard photon kicks the resonance back into the detector even when $M_X < \sqrt{s}e^{-\eta_{\rm max}}$ and distinguishes the signal from VBF backgrounds.

The angular distribution of the ISR photon can be retained for any process $\ell^+\ell^-\rightarrow \gamma \mathcal{F}$ as \cite{Chen:1974wv} 
\begin{align}
    \frac{\sigma_{\ell^+ \ell^- \rightarrow \gamma \mathcal{F}}(s)}{d\cos \theta} &\approx  \frac{1}{2}\int_0^1 d\tau \, \frac{dK(\tau,\,\cos \theta)}{d\cos \theta} \;\hat{\sigma}_{\ell^+ \ell^-\rightarrow \mathcal{F}}(\tau s) ,\\[2ex]
    \frac{dK(\tau,\,\cos \theta)}{d\cos \theta} & \equiv \frac{\alpha}{\pi}\left[ \frac{(1+\tau)^2 + (1-\tau)^2 \cos^2\theta}{(1-\tau)(1-\cos^2\theta)}\right].
\end{align}
Applying an acceptance cut on the photon to be in the rapidity range $[-\eta_{\rm max},\eta_{\rm max}]$ results in the fiducial cross section
\begin{align}
    \sigma_{\ell^+ \ell^- \rightarrow \gamma \mathcal{F}}^{{\rm fid}}(s) &= \int_0^1 d\tau K(\tau,\,\tanh (\eta_{\rm max}))\;\hat{\sigma}_{\ell^+ \ell^-\rightarrow \mathcal{F}}(\tau s),\\[2ex]
    K(\tau,\,\tanh (\eta_{\rm max})) &=\frac{\alpha}{\pi}\left[ \frac{2\eta_{\rm max}\,(1+\tau^2) + (1-\tau)^2 \tanh (\eta_{\rm max})}{1-\tau}\right]
\end{align}
Note that this approaches the leptonic parton luminosity function as $2\eta_{\rm max}\rightarrow \log(Q^2/m_\ell^2)$.

The signal fiducial cross section is obtained by applying the narrow width approximation for $\mathcal{F} = (X\rightarrow \ell' \ell')$
\begin{align}
    \sigma^{(\rm fid)}_{\ell^+ \ell^- \rightarrow \gamma (X\rightarrow \ell'\ell')}(s)  = K(\tau_X,\,\tanh (\eta_{\rm max})\,) \frac{16\pi^2 \mathcal{S}\,\mu_X}{s}.
\end{align}
Final-state leptons may fall outside the detector acceptance, but the recoil from the hard photon limits this effect and we neglect it in the approximate analysis.

The irreducible background comes from replacing the resonance with a $\gamma^*/Z^*$; the cross section is
\begin{align}
    \sigma_{\rm irr,\,ex}(s,\eta_{\rm max}) \approx \left(\frac{4.75 \alpha^2}{s}\right) 2\Delta\,K(\tau_X,\,\tanh (\eta_{\rm max}))
\end{align}
The exclusive search sensitivity is then
\begin{align} \label{eq:Exclusive_Search_Sensitivity}
    \mu_X^{(95\%\, \rm UL)}\approx 10^{-8}\, \left(\frac{\Delta}{0.05}\right)^{1/2}\left(\frac{\sqrt{s}}{10~{\rm TeV}}\right) \left( \frac{L_{\rm int}}{10~{\rm ab}^{-1}}\, K(\tau_X,\,\tanh (\eta_{\rm max}))\right)^{-1/2}.
\end{align}
Results are shown in \cref{fig:sensitivity_MuC}. Compared to the inclusive search, the exclusive sensitivity is degraded by roughly $\left(\log(Q/m_\ell)/\eta_{\rm max}\right)^{1/2}$, but does not suffer the sharp acceptance loss at low masses.

\begin{figure*}[h!]
  \centering

  \subfloat[$e^+e^-$ round]{
    \includegraphics[width=0.48\linewidth]{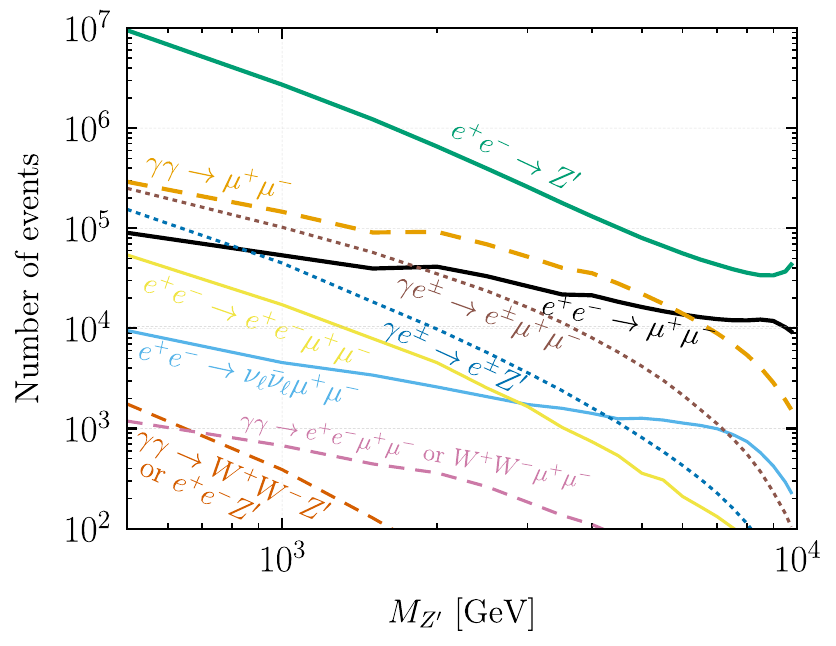}
  }\hfill
  \subfloat[$\gamma\gamma$]{
    \includegraphics[width=0.48\linewidth]{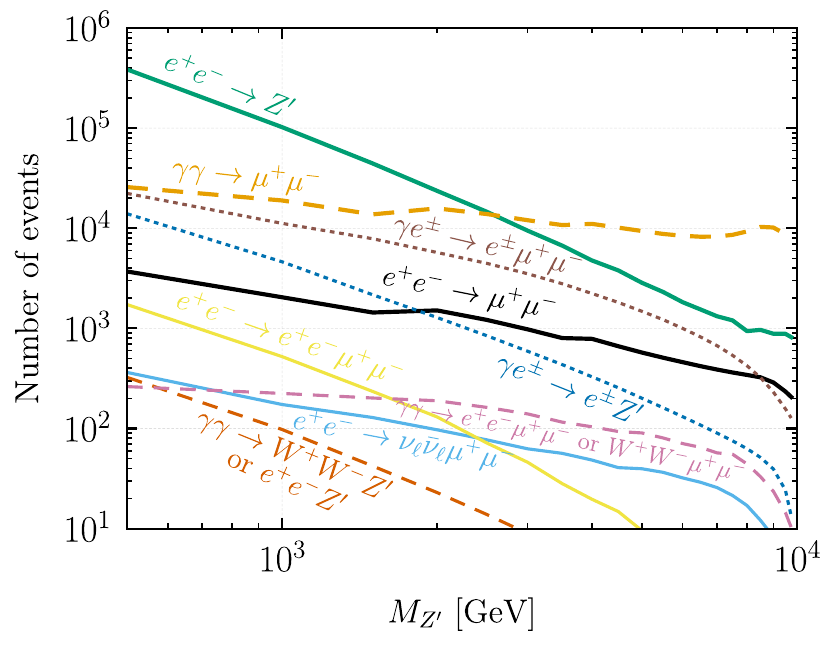}
  }
  \subfloat[$e^-e^-$ flat]{
    \includegraphics[width=0.48\linewidth]{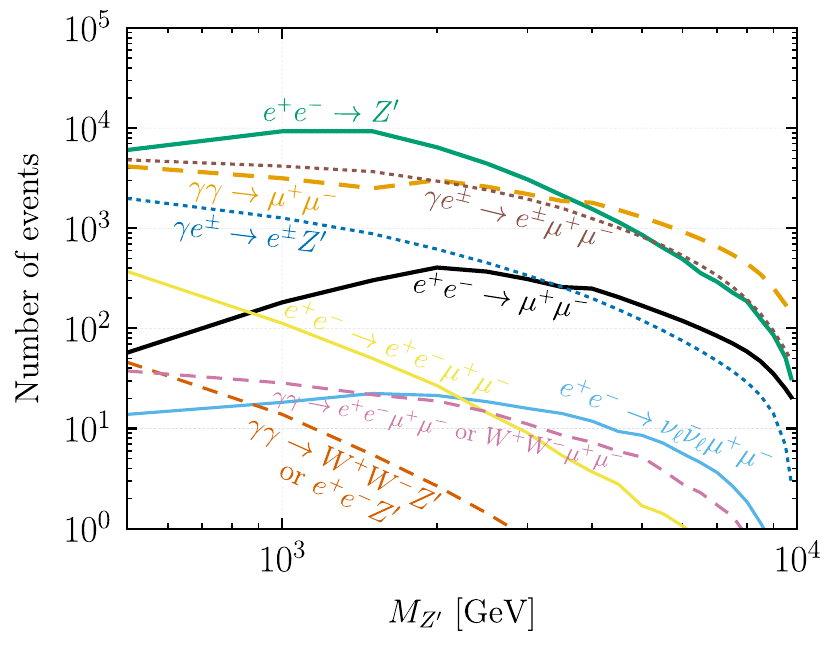}
  }
\caption{Number of events for different processes after cuts in \cref{tab:cuts} for three different WFC configurations: $e^+e^-$ round, $\gamma\gamma$ and $e^-e^-$ flat. We always use ${L}_{\mathrm{geom}}=10\, \mathrm{ab}^{-1}$. The signal assumes $\varepsilon=0.1$ while the linestyles indicate the initial state: $e^+e^-$ (solid), $\gamma\gamma$ (dashed) and $e^\pm \gamma$ (dotted). }
    \label{fig:All_PWFA_events}
\end{figure*}

\section{Additional Analysis Details} \label{app:analysis-details}

\subsection{Event Yields and Additional Backgrounds}

In the main text we considered only the most relevant processes; here we further discuss subdominant processes which contribute to searches as well as processes which become important for $e^- e^-$ configurations. The expected number of events, after acceptance and selection cuts described in \cref{tab:cuts}, for several values of $M_{Z'}$ are shown for a MuC in \cref{tab:MuC_events} and for the different WFC configurations in \cref{tab:PWFA_events}.  

The fiducial cross section as a function of $M_{Z'}$ for different processes at a muon collider is shown in \cref{fig:LEPDFvsAnalytical}. We also show a plot of the number of events passing cuts for three WFC configurations in \cref{fig:All_PWFA_events} as a function of $M_{Z'}$. We see $e^-e^-$ WFC configurations have a background from $\gamma e^- \to e^-\mu^+\mu^-$ (where the final state $e^-$ is lost down the beam-line) which is relevant for the inclusive search. The signal and irreducible background for this WFC configuration also have different shapes relative to other configurations because the rapidity distribution of events with $e^-e^-$ flat configuration is not as broad as in the other cases.

\begin{table*}[h!]
\centering
\caption{ Estimated number of detected events in MuC searches, using $\sqrt{s}=\SI{10}{\TeV}$ and $L=10\,\mathrm{ab}^{-1}$. We show several values of $M_{Z'}$ with effects of selection cuts on invariant mass and rapidity used in the analysis. The signal is evaluated at $\varepsilon=0.1$. Masses are in GeV and we consider only $Z'\to e^+e^-$ decays.}
\label{tab:MuC_events}

{\renewcommand{\arraystretch}{1.3}
\resizebox{\textwidth}{!}{%
\begin{tabular}{c c c c c c c c}
\toprule
$M_{Z'}$ & Cuts & $\mu^+\mu^-\to Z'$& $\mu^+\mu^-\to e^+ e^-$& $\mu^+\mu^-\to \mu^+\mu^-e^+ e^-$& $\mu^+\mu^-\to \nu_\ell\bar{\nu}_\ell e^+ e^-$& $\mu^+\mu^-\to Z'\gamma$ & $\mu^+\mu^-\to e^+ e^-\gamma$\\
\hline
\multirow{2}{*}{1000} & $990<M_{e^+e^-}<1010$& 232& 7.43& 1670 & 66.0& 88.9& 2.73\\ 
    & $\&\,2.292<|y_{e^+e^-}|<2.312$ & 102& 2.11& 0.496 & 0.049& -& - \\ 
\hline
\multirow{2}{*}{2500} & $2475<M_{e^+e^-}<2525$& 551& 20.4& 150& 19.9& 105& 4.54\\ 
        & $\&\,1.376<|y_{e^+e^-}|<1.396$& 401& 9.96& 1.17& 0.224& -& - \\ 
\hline
\multirow{2}{*}{8000} & $7920<M_{e^+e^-}<8080$& 2190& 75.3& 1.27& 53.4& 502& 27.5\\ 
        & $\&\,0.213<|y_{e^+e^-}|<0.223$& 1800& 43.7& 0.211& 1.53& -& - \\ 
\bottomrule
\end{tabular}}%
} \end{table*}

\begin{table*}[h!]
\centering
\caption{ Estimated number of detected events in the WFC search, using $L_{\mathrm{geom}}=10\,\mathrm{ab}^{-1}$. We show several values of $M_{Z'}$ with the effect of the invariant mass selection cut. The signal is evaluated at $\varepsilon=0.1$. Masses are in GeV. ``Other backgrounds'' accounts for $e^+e^-\to e^+e^-\mu^+\mu^-$, $e^+e^-\to \nu_\ell\bar{\nu}_\ell\mu^+\mu^-$, $e^\pm\gamma\to e^\pm\mu^+\mu^-$, $\gamma\gamma\to e^+e^-\mu^+\mu^-$, $\gamma\gamma\to W^+W^-\mu^+\mu^-$. ``Other signals'' accounts for $e^\pm\gamma\to e^\pm Z'$, $\gamma\gamma\to W^+W^-Z'$, $\gamma\gamma\to e^+e^-Z'$}
\label{tab:PWFA_events}
{\renewcommand{\arraystretch}{1.3}
\resizebox{\textwidth}{!}{\begin{tabular}{c c c c c c c c c}
\toprule
\textbf{Configuration} & $M_{Z'}$ & min $M_{\mu^+\mu^-}$ & max $M_{\mu^+\mu^-}$ & $e^+e^- \to Z' \to \mu^+ \mu^-$ & $e^+e^- \to \mu^+ \mu^-$ & $\gamma\gamma \to \mu^+ \mu^-$ & Other backgrounds & Other signals \\
\hline

\multirow{3}{*}{$e^+e^-$ round} & 1000 & 985 & 1015 & $2.72\times10^6$& $5.37\times10^4$& $1.46\times10^5$& $1.25\times10^5$& $4.54\times10^4$\\ 
& 2500 & 2406 & 2594 & $3.95\times10^5$& $3.32\times10^4$& $6.91\times10^4$& $2.86\times10^4$& 5780 \\
& 8000 & 7040 & 8960 & $3.60\times10^4$& $1.20\times10^4$& 5430& 1390& 112 \\
\hline

\multirow{3}{*}{$e^+e^-$ flat} & 1000 & 985 & 1015 & $1.09\times10^5$& 2160 & $1.76\times10^4$& $2.70\times10^4$& 5790\\ 
& 2500 & 2406 & 2594 & $4.31\times10^4$& 3600& $1.34\times10^4$& 9230& 1530 \\
& 8000 & 7040 & 8960 & $1.63\times10^4$& 5450& 2080& 907& 72.5 \\
\hline

\multirow{3}{*}{$\gamma\gamma$} & 1000 & 985 & 1015 & $1.02\times10^5$& 2050 & $1.90\times10^4$& $1.21\times10^4$ & 4720\\ 
& 2500 & 2406 & 2594 & $1.46\times10^4$& 1215& $1.39\times10^4$& 4770& 858 \\
& 8000 & 7040 & 8960 & 939& 343& 9300& 487& 63.2 \\
\hline

\multirow{3}{*}{$e^-e^-$ round} & 1000 & 985 & 1015 & $3.71\times10^4$& 727 & 7230& 7410 & 2430\\ 
& 2500 & 2406 & 2594 & $1.20\times10^4$& 1020& 5590& 3660& 695 \\
& 8000 & 7040 & 8960 & 352& 121& 914& 266& 38.5 \\
\hline

\multirow{3}{*}{$e^-e^-$ flat} & 1000 & 985 & 1015 & 9310& 182 & 3160& 4340 & 1280\\ 
& 2500 & 2406 & 2594 & 4440& 369& 2600& 2460& 454 \\
& 8000 & 7040 & 8960 & 186& 58.7& 444& 198& 29.0 \\
\bottomrule

\end{tabular}}}
\end{table*}
\subsection{Photon-Photon Collisions at Wakefield Colliders}
\begin{figure*}[h!]
    \centering
    \includegraphics[height=0.4\linewidth]{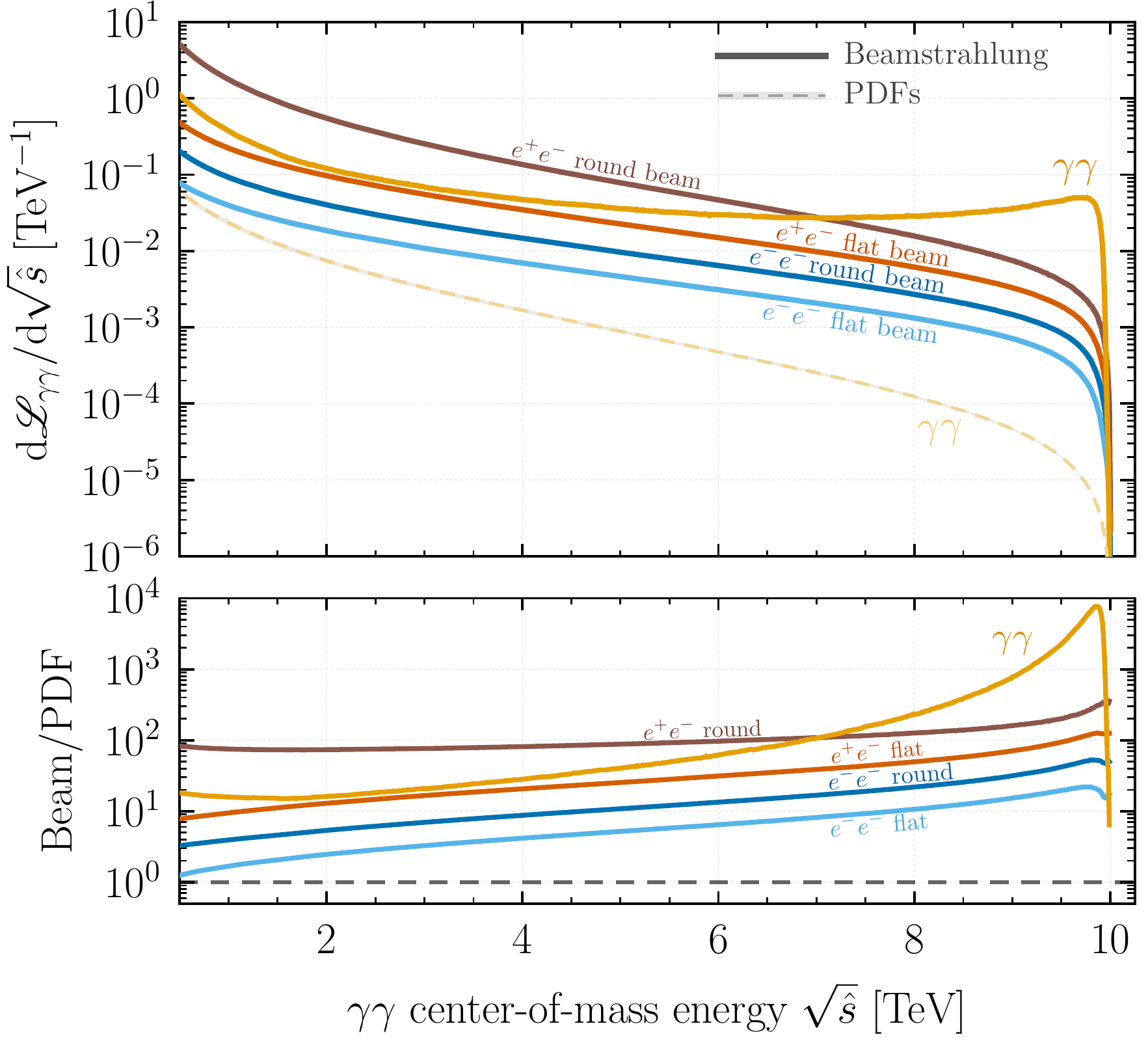}\hspace{1cm}
    \raisebox{0.6ex}{\includegraphics[height=0.4\linewidth]{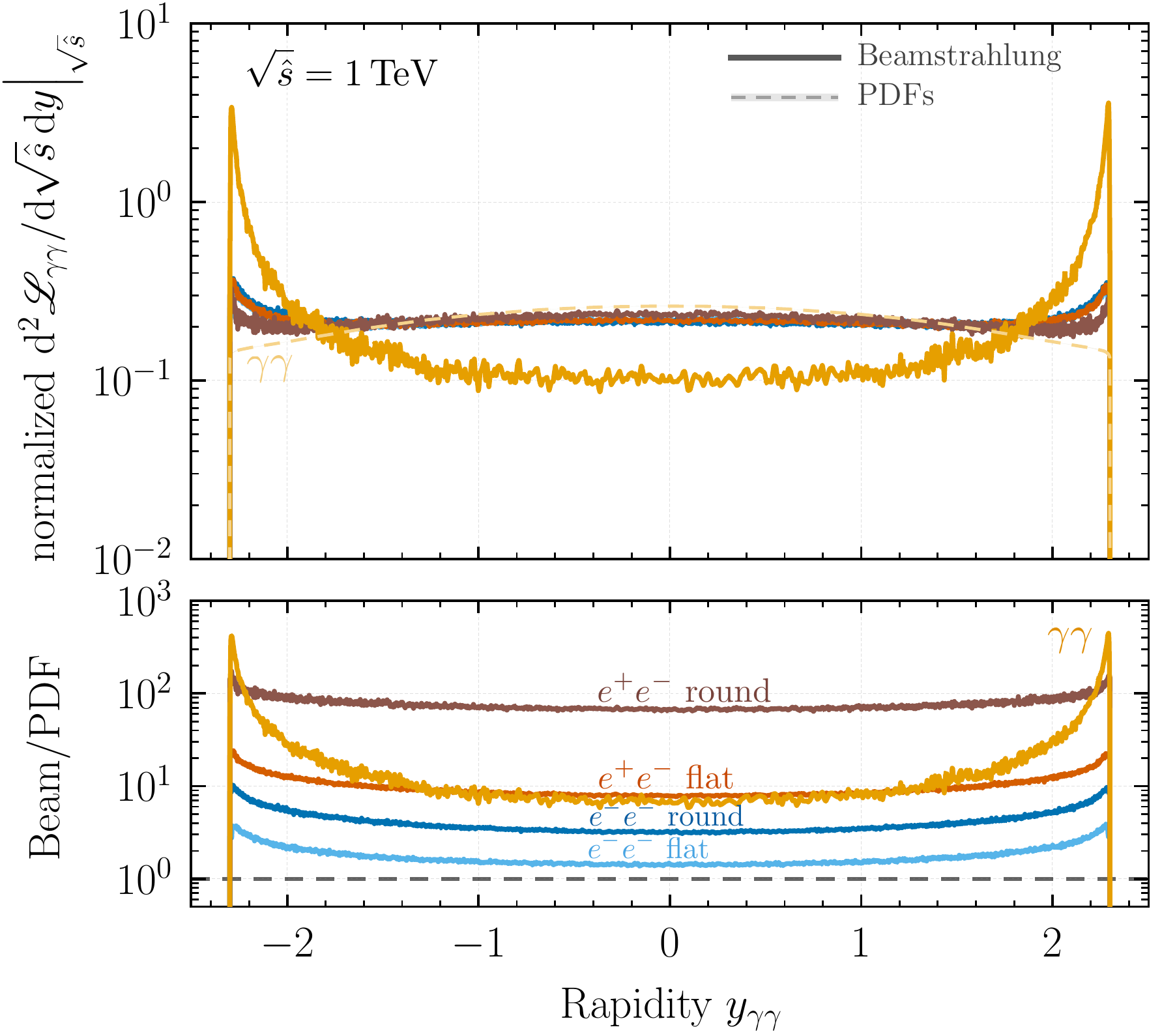}}
    \caption{ Luminosity as a function of $\sqrt{\hat{s}}$ for $\gamma\gamma$ initial states. Details in this figure are the same as \cref{fig:beam-beam-vs-machine}.}
    \label{fig:beam-beam-vs-machine-gg}
\end{figure*}

As emphasized in the main text, beamstrahlung creates broad spectra of secondary particles, offering additional physics potential. \cref{fig:beam-beam-vs-machine-gg} shows luminosity spectra for $\gamma \gamma$ collisions for five WFC configurations. $\gamma \gamma$ colliders provide high luminosity at the maximum energy, while all WFC configurations maintain sizable luminosity at lower energies.
\vspace{-0.75cm}
\subsection{Rapidity Distributions}
\begin{figure*}[h!]
  \centering

  \subfloat[Normalized rapidity distributions\label{fig:Rapidity_comparison_norm}]{
    \includegraphics[width=0.45\linewidth]{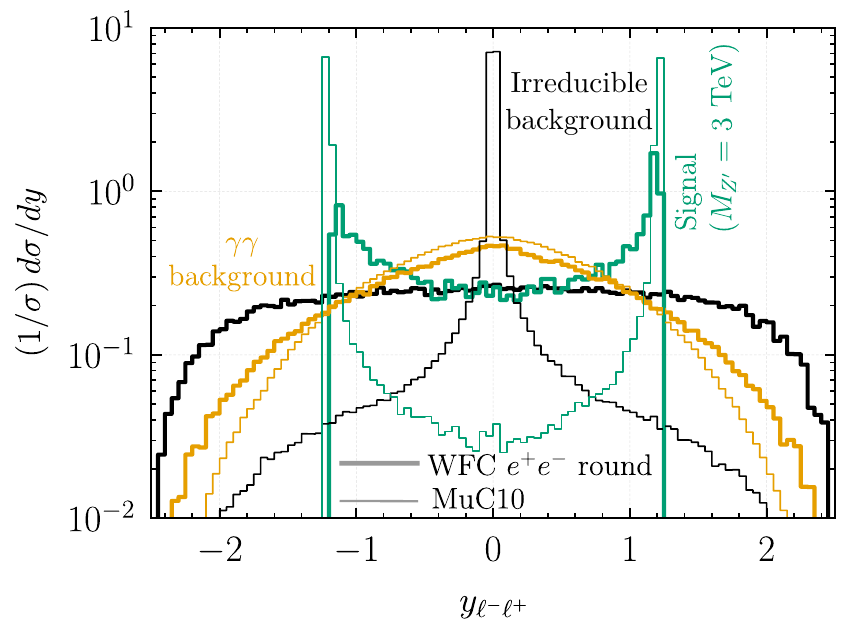}
  }\hfill
  \subfloat[Unnormalized event counts (with mass cut)\label{fig:Rapidity_comparison_N_cut}]{
    \includegraphics[width=0.45\linewidth]{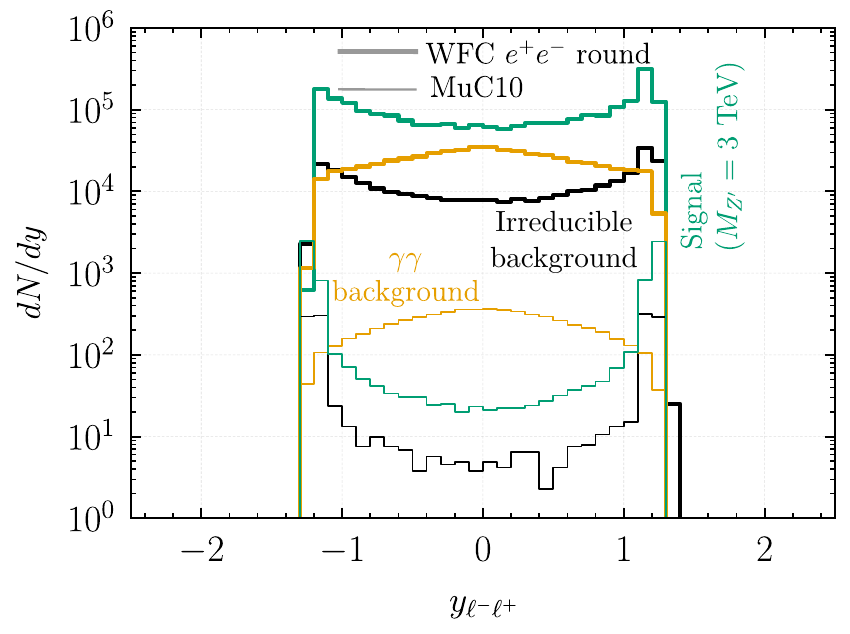}
  }
\caption{
    Rapidity distributions for inclusive search processes at a 10 TeV MuC and $e^+e^-$ round WFC. \textbf{(a)} Signal ($M_{Z'}=\SI{3}{\TeV}$, $\varepsilon=0.1$) and background distributions with no selection cuts applied. WFC events include collision energies above 500 GeV. \textbf{(b)} Event yields assuming $L_{\rm geom}=10\,\mathrm{ab}^{-1}$ for both colliders, after applying the invariant mass selection cut from \cref{tab:cuts}.
    }
    \label{fig:Rapidity_Comparison_Combined}
\end{figure*}

Rapidity distributions are crucial for resonance searches at both a MuC and WFC. \cref{fig:Rapidity_Comparison_Combined} shows example distributions for a MuC and an $e^+e^-$ WFC with round beams. 
The WFC distribution is broader, whereas the MuC distribution peaks at $|y|\approx y_{Z'}$. 
This difference accounts for both the weakness of the WFC's di-lepton rapidity cut and the MuC's lost sensitivity in inclusive searches for $M_{Z'}<\SI{1}{\TeV}$ (when the di-lepton escapes down the beam pipe).
\vspace{-0.5cm}

\end{document}